\documentclass[preprintnumbers,floatfix,letterpaper,aps,prd,epsfig,nofootinbib]{revtex4-1}
\pdfoutput=1
\usepackage{bm,graphicx,dcolumn,epstopdf,epsf, latexsym,mathbbol, amssymb,amsmath,color,slashed, mathrsfs,mathcomp,simplewick}
\usepackage[center]{subfigure}
\usepackage{multirow}
\usepackage{makecell}
\usepackage[colorlinks,linkcolor=blue,citecolor=blue,urlcolor=blue]{hyperref}

\begin{document}
\allowdisplaybreaks
 \newcommand{\bq}{\begin{equation}}
 \newcommand{\eq}{\end{equation}}
 \newcommand{\bqn}{\begin{eqnarray}}
 \newcommand{\eqn}{\end{eqnarray}}
 \newcommand{\nb}{\nonumber}
 \newcommand{\lb}{\label}
 \newcommand{\f}{\frac}
 \newcommand{\p}{\partial}
\newcommand{\PRL}{Phys. Rev. Lett.}
\newcommand{\PLB}{Phys. Lett. B}
\newcommand{\PRD}{Phys. Rev. D}
\newcommand{\CQG}{Class. Quantum Grav.}
\newcommand{\JCAP}{J. Cosmol. Astropart. Phys.}
\newcommand{\JHEP}{J. High. Energy. Phys.}
\newcommand{\red}{\textcolor{black}}

\title{Post-Newtonian parameters of ghost-free parity-violating gravities}

\author{Jin Qiao${}^{a,b,c}$}
\email{qiaojin@zjut.edu.cn}

\author{Tao Zhu${}^{d,e}$}
\email{zhut05@zjut.edu.cn}

\author{Guoliang Li${}^{a,b}$}
\email{guoliang@pmo.ac.cn}

\author{Wen Zhao${}^{b,c}$}
\email{wzhao7@ustc.edu.cn}

\affiliation{
${}^{a}$  Purple Mountain Observatory, Chinese Academy of Sciences, Nanjing, 210023, P.R.China\\
${}^{b}$ School of Astronomy and Space Sciences, University of Science and Technology of China, Hefei, 230026, P.R.China\\
${}^{c}$ Department of Astronomy, University of Science and Technology of China, Hefei 230026, P.R.China \\
${}^{d}$ Institute for theoretical physics and Cosmology, Zhejiang University of Technology, Hangzhou, 310032, P.R.China\\
${}^{e}$ United center for gravitational wave physics (UCGWP), Zhejiang University of Technology, Hangzhou, 310032, P.R.China; \\
}
\date{\today}

\begin{abstract}
We investigate the slow-motion and weak-field approximation of the general ghost-free parity-violating (PV) theory of gravity in the parametrized post-Newtonian (PPN) framework and derive the perturbative field equations, which are modified by the PV terms of this theory. The complete PPN parameters are obtained by solving the perturbative field equations. We find that all the PPN parameters are exactly the same as those in general relativity, except for an extra parameter $\kappa$, which is caused by the new \red{curl-type} term in the gravitomagnetic sector of the metric in this theory. We calculate the precession effects of gyroscopes in this theory and constrain the model parameters by the observations of the Gravity Probe B experiment.
\end{abstract}

\maketitle

\section{Introduction}
\renewcommand{\theequation}{1.\arabic{equation}} \setcounter{equation}{0}

Tests of gravity have been widely concerned since Einstein first formulated general relativity (GR) \cite{Clifton:2011jh,Sotiriou:2008rp,Nojiri:2017ncd,Capozziello:2011et}. At present, the results of almost all gravity testing experiments show that GR is the most successful theory of gravity among many gravitational theories. Nevertheless, there are other theories of gravity which also satisfy the various high precision experimental constraints and remain candidates for the alternative theory of gravity \cite{Nojiri:2010wj}. Moreover, GR still faces difficulties in both theoretically (e.g. singularity, quantization, etc) and observationally (e.g. dark matter, dark energy, etc). Modified gravity is considered to be one of the effective ways to solve these anomalies \cite{Cognola:2006eg,Copeland:2006wr,Frieman:2008sn,Li:2011sd}. Therefore, the tests of the modified gravities are essential to confirm the final theory of gravity.

In this article, we focus on the parity-violating (PV) gravities, which are a class of alternative theories of gravity that the gravitational action is modified by including the PV terms. Parity symmetry implies that a directional flipping to the left and right does not change the laws of physics. It is well known that nature is parity-violating since the first discovery of parity violation in weak interactions \cite{LeeYang}. Although in GR, the parity symmetry is conserved, many PV gravities with different action forms have been proposed for different motivations \cite{23,Yunes:2010yf,25,Alexander:2017jmt,28,29,PHor:2009,31,32,33,Li:2021mdp}. PV theories of gravity have been studied in the cosmology, as well as in gravitational waves \cite{Wang:2012fi,Zhu:2013fja,Takahashi:2009wc,Maldacena:2011nz,Yagi:2017zhb,Gao:2019liu,Zhao:2019szi,cs1,Alexander:2017jmt,Yunes:2010yf,Nishizawa:2018srh,Li:2021mdp,Li:2021wij,Li:2020xjt,Hou:2021oxe,Wu:2021ndf,Gong:2021jgg}. Gravitational parity violation can produce the birefringence effect of (primordial) gravitational waves, where the modified dispersion relation can lead to velocity birefringence \cite{PHor:2009} and the altered friction can cause amplitude birefringence of gravitational waves \cite{Lue:1998mq, 23}. The imprints of these birefringence effects can be detected by the laser interference gravitational wave detectors \cite{Yagi:2017zhb,WYF2021a,WYF2021b,Wang:2020pgu,HQ2021,shao} and/or by the polarization of cosmic microwave background radiations \cite{Lue:1998mq,Saito:2007kt,Wang:2012fi,Qiao:2019hkz}.

In addition, the external environment of compact bodies such as binary pulsars or solar system objects \cite{Smith:2007jm,CS_E,Stairs:2003eg,Manchester:2015mda,Kramer:2016kwa}, also provides the most accessible testbed to test the PV theories of gravity. The parameterized post-Newtonian (PPN) approximation \cite{will2014,will2018th} is successful and extensively used to test the slow-motion and weak-field limit of the gravity theories \cite{Stephon2007,Flathmann:2020zyj,Zhang:2016njn,Lin:2013tua,Foster:2005dk,Rao:2021azn}, including PV theories of gravity. For a specific Chern-Simons (CS) gravity, Refs.\cite{Stephon2007a,Stephon2007} have calculated the PPN parameters and shown that the vectorial sector of the metric is modified by a new \red{curl-type} term, leading to a correction to the Lense-Thirring effect.
Other PV theories of gravity, for example, Ho$\rm \check{r}$ava-Lifshitz gravity \cite{Lin:2013tua},  teleparallel gravity \cite{Rao:2021azn}, etc, also have been examined in the slow-motion and weak-field limit of the system by using the PPN approximation.

Recently, based on the specific parity-violating CS modified gravity, a ghost-free parity-violating theory of gravity has been explored in Ref. \cite{Crisostomi:2017ugk} by including higher derivatives of the coupling scalar field. In this theory of gravity, we have studied the propagation of gravitational waves and found that both amplitude and velocity birefringence exist \cite{TZ2019,Zhao:2019xmm}.  We also investigated circular polarization of the primordial gravitational waves in this gravity and obtained a nonzero degree of circular polarization \cite{Qiao:2019hkz}.

As an extension of these works, in this paper, we will calculate the PPN parameters of the ghost-free PV theory of gravity to test whether it is compatible with the solar system experiments. We use the PPN approximation to expand the modified field equations. By solving the perturbative field equations, we get the full PPN metric and the PPN parameters of this theory. We find that this metric contains an extra \red{curl-type} term, which is similar to in CS theory \cite{Stephon2007a,Stephon2007} except for the coupling parameter. In order to constrain this theory, we calculate the modification to the rate of angular precession in a specific experiment frame and find that the modification contributes in two directions of the geodetic effect of gyroscopes {in a complete circular satellite orbital motion period.
However, in the period of a complete Earth orbital period, the contributions in these two directions include the oscillation terms produced by the Earth's orbital motion. We ignore these oscillation terms in a complete Earth's orbital cycle, the modification to the rate of angular precession only contributes in one direction, which exactly corresponds to the changes of GR's frame-dragging precession.
This allows us to apply the observation of the Gravity Probe B (GPB) experiment to constrain the model parameters}.

This paper is organized as follows. In Sec.\ref{sec2}, we briefly introduce the theory of ghost-free PV gravity and the modified field equations.
In Sec.\ref{sec3}, we describe the basics of the PPN framework and expand the modified field equations to the required order in the metric perturbation.
In Sec.\ref{sec4}, we solve the equations in the PN approximation to obtain the PPN parameters.
In Sec.\ref{sec5}, we constrain this theory with the frame-dragging effect. Summary and discussions are given in Sec.\ref{sec6}.

Throughout this paper, the metric convention is chosen as $(-, +, +, +)$, and greek indices $(\mu, \nu, \cdot\cdot\cdot)$ run over $0, 1, 2, 3$ and the latin indices $(i, j, k, \cdot\cdot\cdot)$ run over $1, 2, 3$. We set the units to $c = \hbar=1$.

\section{Ghost-free parity-violating gravities\label{sec2}}
\renewcommand{\theequation}{2.\arabic{equation}} \setcounter{equation}{0}

The action of general PV gravity can be written in the following form
\bqn\lb{action}
S &=& \frac{1}{16\pi G} \int d^4 x \sqrt{-g}(R+\mathcal{L}_{\rm PV})  + \int d^4 x \sqrt{-g} ( \mathcal{L}_{\phi} + \mathcal{L}_{\rm other}),
\eqn
where $R$ is the Ricci scalar, $\mathcal{L}_{\rm PV}$ is a parity-violating Lagrangian, $\mathcal{L}_\phi$ is the Lagrangian for a scalar field, which may be coupled non-minimally to gravity, and $\mathcal{L}_{\rm other}$ denotes other matter fields. As one of the simplest examples, we consider the action of the scalar field
\bqn
\mathcal{L}_\phi =   \frac{1}{2} g^{\mu \nu} \partial_\mu \phi \partial_\nu \phi +V(\phi).
\eqn
Here $V(\phi)$ denotes the potential of the scalar field. The parity-violating Lagrangian $\mathcal{L}_{\rm PV}$ has different expressions for different theories. CS modified gravity with Pontryagin term coupled with a scalar field is a widely studied PV gravity in the previous works. The Lagrangian of CS reads \cite{Alexander:2009tp}
\bqn
\mathcal{L}_{\rm CS} = \frac{1}{4}\vartheta(\phi) \;^*R R,
\eqn
where
\bqn
\;^*R R=\frac{1}{2} \varepsilon^{\mu\nu\rho\sigma} R_{\rho\sigma \alpha\beta} R^{\alpha \beta}_{\;\;\;\; \mu\nu}
\eqn
is the Pontryagin density with $\varepsilon^{\rho \sigma \alpha \beta}$ the Levi-Civit\'{a} tensor defined in terms of the antisymmetric symbol $\epsilon^{\rho \sigma \alpha \beta}$ as $\varepsilon^{\rho \sigma \alpha \beta}=\epsilon^{\rho \sigma \alpha \beta}/\sqrt{-g}$ and the CS coupling coefficient $\vartheta(\phi)$ being an arbitrary function of $\phi$. CS modified gravity is an effective extension of GR that captures leading-order, gravitational parity-violating term. The similar versions of this theory were suggested in the context of string theory \cite{string,string1}, and three-dimensional topological massive gravity \cite{massive}. However, this theory has higher-derivative field equation, which induces the dangerous Ostrogradsky ghosts. For this reason, CS modified gravity can only be treated as a low-energy truncation of a fundamental theory. To cure this problem, the extension of CS gravity by considering the terms which involve the derivatives of a scalar field is recently proposed in \cite{Crisostomi:2017ugk}.
\red{The action is generalised in this paper by including first and second derivatives of the scalar field: $\phi_\mu \equiv \partial_\mu \phi$ and $\phi_{\sigma\nu} \equiv \nabla_\sigma \phi_\nu $. }

 $\mathcal{L}_{\rm PV1}$ is the Lagrangian containing the first derivative of the scalar field, which is given by
\bqn\lb{LMPV1}
\mathcal{L}_{\rm PV1} &=& \sum_{\rm A=1}^4  a_{\rm A}(\phi, \phi^\mu \phi_\mu) L_{\rm A},\label{lv1}
\eqn
where
\bqn
L_1 &=& \varepsilon^{\mu\nu\alpha \beta} R_{\alpha \beta \rho \sigma} R_{\mu \nu}{}^{\rho}{}_{\lambda} \phi^\sigma \phi^\lambda,\nonumber\\
L_2 &=&  \varepsilon^{\mu\nu\alpha \beta} R_{\alpha \beta \rho \sigma} R_{\mu \lambda }^{\; \; \;\rho \sigma} \phi_\nu \phi^\lambda,\nonumber\\
L_3 &=& \varepsilon^{\mu\nu\alpha \beta} R_{\alpha \beta \rho \sigma} R^{\sigma}_{\;\; \nu} \phi^\rho \phi_\mu,\nonumber\\
L_4 &=&  \varepsilon^{\mu\nu\rho\sigma} R_{\rho\sigma \alpha\beta} R^{\alpha \beta}_{\;\;\;\; \mu\nu} \phi^\lambda \phi_\lambda,\nonumber
\eqn
with $\phi^\mu \equiv \nabla^\mu \phi$, and $a_{\rm A}$ are a priori arbitrary functions of $\phi$ and $\phi^\mu \phi_\mu$. In order to avoid the Ostrogradsky modes in the unitary gauge (where the scalar field depends on time only), it is required that $4a_1+2 a_2+a_3 +8 a_4=0$. With this condition, the Lagrangian in Eq.(\ref{lv1}) does not have any higher order time derivative of the metric, but only higher order space derivatives.

One can also consider the terms which contain second derivatives of the scalar field. Focusing on only these that are linear in Riemann tensor and linear/quadratically in the second derivative of $\phi$, the most general Lagrangian $\mathcal{L}_{\rm PV2}$ is given by \cite{Crisostomi:2017ugk}
\bqn\lb{LMPV2}
\mathcal{L}_{\rm PV2} &=& \sum_{\rm A=1}^7 b_{\rm A} (\phi,\phi^\lambda \phi_\lambda) M_{\rm A},
\eqn
where
\bqn
M_1 &=& \varepsilon^{\mu\nu \alpha \beta} R_{\alpha \beta \rho\sigma} \phi^\rho \phi_\mu \phi^\sigma_\nu,\nonumber\\
M_2 &=& \varepsilon^{\mu\nu \alpha \beta} R_{\alpha \beta \rho\sigma} \phi^\rho_\mu \phi^\sigma_\nu, \nonumber\\
M_3 &=& \varepsilon^{\mu\nu \alpha \beta} R_{\alpha \beta \rho\sigma} \phi^\sigma \phi^\rho_\mu \phi^\lambda_\nu \phi_\lambda, \nonumber\\
M_4 &=& \varepsilon^{\mu\nu \alpha \beta} R_{\alpha \beta \rho\sigma} \phi_\nu \phi_\mu^\rho \phi^\sigma_\lambda \phi^\lambda, \nonumber\\
M_5 &=& \varepsilon^{\mu\nu \alpha \beta} R_{\alpha \rho\sigma \lambda } \phi^\rho \phi_\beta \phi^\sigma_\mu \phi^\lambda_\nu, \nonumber\\
M_6 &=& \varepsilon^{\mu\nu \alpha \beta} R_{\beta \gamma} \phi_\alpha \phi^\gamma_\mu \phi^\lambda_\nu \phi_\lambda, \nonumber\\
M_7 &=& (\nabla^2 \phi) M_1,\nonumber
\eqn
with $\phi^{\sigma}_\nu \equiv \nabla^\sigma \nabla_\nu \phi$. Similarly, in order to avoid the Ostrogradsky modes in the unitary gauge, the following conditions should be imposed: $b_7=0$, $b_6=2(b_4+b_5)$ and $b_2=-A_*^2(b_3-b_4)/2$, where $A_*\equiv \dot{\phi}(t)/N$ and $N$ is the lapse function. In this paper, we consider a general scalar-tensor theory with parity violation, which contains all the terms mentioned above. So, the parity-violating term in Eq.(\ref{action}) is given by
\bqn
\mathcal{L}_{\rm PV} = \mathcal{L}_{\rm CS} + \mathcal{L}_{\rm PV1} + \mathcal{L}_{\rm PV2}.
\eqn
Therefore, the CS modified gravity in \cite{Alexander:2009tp}, and the ghost-free parity-violating gravities discussed in \cite{Crisostomi:2017ugk} are all the specific cases of this Lagrangian. The coefficients $\vartheta$, $a_{\rm A}$ and $b_{\rm A}$ depend on the scalar field $\phi$ and its evolution.

\red{In principle, we can also continue to construct the higher-order scalar field derivatives coupled with the curvatures, such as $\mathcal{L}_{\rm PV3} $, $\mathcal{L}_{\rm PV4} $, etc. However, the contributions of these higher-order coupling terms are expected to be at the higher-order in the perturbative expansions of the PPN approximation. From the calculations for $\mathcal{L}_{\rm PV1} $ and $\mathcal{L}_{\rm PV2} $, it can be observed that $\mathcal{L}_{\rm PV2} $ does not contribute to any PPN parameters at the leading order. Therefore, it is sufficient here that we only consider the action with the first and the second derivative terms of the scalar field as in \cite{Crisostomi:2017ugk}.}

Variation of the action with respect to the metric tensor $g_{\mu\nu}$,  one obtains the field equation of the theory, which is,
\bqn\lb{2.8}
R_{\mu\nu}- \frac{1}{2} g_{\mu\nu} R + C_{\mu\nu} +  A_{\mu\nu} +  B_{\mu\nu} =8 \pi G (T_{\mu\nu}^{\rm m} + T_{\mu\nu}^{\phi}),\lb{field}
\eqn
where
\bqn
C_{\mu\nu} \equiv \frac{1}{\sqrt{-g}} \frac{\delta (\sqrt{-g} \mathcal{L}_{\rm CS})}{\delta g^{\mu\nu}}= \;^*R^\beta{}_{\mu\nu}{}^\alpha \nabla_\alpha\nabla_\beta \theta + (\nabla_\alpha \theta) \epsilon^{\alpha \beta \gamma}_\mu \nabla_\gamma R_{\nu \beta }, \\
A_{\mu\nu} \equiv \frac{1}{\sqrt{-g}} \frac{\delta (\sqrt{-g} \mathcal{L}_{\rm PV1})}{\delta g^{\mu\nu}}, \;\;\; B_{\mu\nu} \equiv \frac{1}{\sqrt{-g}} \frac{\delta (\sqrt{-g} \mathcal{L}_{\rm PV2})}{\delta g^{\mu\nu}},\\
T_{\mu\nu}^{\rm m} = - \frac{2}{\sqrt{-g}} \frac{\delta (\sqrt{-g} \mathcal{L}_{\rm m})}{\delta g^{\mu\nu}}, \;\;\; T_{\mu\nu}^{\phi} = - \frac{2}{\sqrt{-g}} \frac{\delta (\sqrt{-g} \mathcal{L}_{\phi })}{\delta g^{\mu\nu}}.
\eqn
The expressions of $A_{\mu\nu}$, $B_{\mu\nu}$, $T_{\mu\nu}^\phi$ are given in Appendix A. The equation of the scalar field can be obtained by varying the action (\ref{action}) with respect to the scalar field $\phi$, which gives
\bqn\lb{2.12}
-16\pi G \nabla^2 \phi + 16\pi G V_{,\phi}(\phi) + \vartheta_{,\phi} \;^*R R + F_{,\phi} =0,
\eqn
where $F_{,\phi}$ is given in the Appendix A.

For the field equation (\ref{field}), it is convenient to write it into a more suitable form. Taking the trace of (\ref{field}) one obtains
\bqn
R =  C+A+B-8\pi G (T^{\rm m} + T^{\phi}), \lb{RR}
\eqn
where $C\equiv g^{\mu\nu} C_{\mu\nu}$, $A\equiv g^{\mu\nu} A_{\mu\nu}$, $B\equiv g^{\mu\nu} B_{\mu\nu}$, $T^{\rm m} \equiv g^{\mu\nu} T^{\rm m}_{\mu\nu}$, and $T^\phi \equiv g^{\mu\nu} T^\phi_{\mu\nu}$. Replacing Eq.~(\ref{RR}) back in eq. (\ref{field}), the latter becomes,
\bqn
R_{\mu\nu} + C_{\mu\nu} + A_{\mu\nu} + B_{\mu\nu} &=& 8\pi G (C + A + B) g_{\mu\nu} - 4\pi G( T^m + T^\phi) g_{\mu\nu}
   + 8\pi G (T^m_{\mu\nu} + T^\phi_{\mu\nu}),
\eqn
where the trace of these tensor $C=g^{\mu\nu} C_{\mu\nu}$, $A=g^{\mu\nu} A_{\mu\nu}$ and $B=g^{\mu\nu} B_{\mu\nu}$ vanishes identically. The field equation reduces to
\bqn\lb{feq}
R_{\mu\nu} + C_{\mu\nu} + A_{\mu\nu} + B_{\mu\nu}  &=& 8\pi G \left[T^m_{\mu\nu} - \frac{1}{2}g_{\mu\nu}T^m + \left( T^\phi_{\mu\nu}  - \frac{1}{2}g_{\mu\nu}T^\phi \right) \right] \nb\\
    &=& 8\pi G \left(S^m_{\mu\nu}  + S^\phi_{\mu\nu}  \right) ,
\eqn
where $S_{\mu\nu}=T_{\mu\nu} - \frac{1}{2}g_{\mu\nu}T$. In order to validate these field equations, in Appendix B, we expand them in the weak gravitational fields, and derive the propagation field of gravitational wave in this theory, which is consistent with the corresponding results in the previous works \cite{TZ2019,Qiao:2019hkz}.

\section{Parametrized Post-Newtonian Expansion \label{sec3}}
\renewcommand{\theequation}{3.\arabic{equation}} \setcounter{equation}{0}

In this section, we start to consider the PPN formalism of the ghost-free PV gravities.  One of important assumption of the PPN formalism is that the matter which acts as the source of the gravitational field is given by a perfect fluid. The velocity of the matter in a particular, fixed frame of reference is small, measured in units of the speed of light, and that all physical quantities relevant for the solution of the gravitational field equations can be expanded in orders of this velocity. In this section, we discuss how this expansion in velocity orders proceeds for the quantities we need in our calculation in the following sections, in particular for parity-violating terms.

The energy-momentum tensor of the matter field thus can be expressed in the perfect fluid form, which is
\bqn
T^{ \mu\nu}_{\rm m} = (\rho+\rho \Pi + p) u^\mu u^\nu+p g^{\mu\nu},
\eqn
where $\rho$, $\Pi$, $p$, and $u^\mu$ are the rest energy density, specific internal energy, pressure, and the four velocity of the matter field respectively. The four velocity satisfies the normalization condition $u^\mu u_{\nu} = -1$. In the PPN formalism \cite{will2014}, we focus on slow motion. So, the last three components of \red{the velocity $v^i \equiv dx^i/dt=u^i/u^t$ }are essentially small. To compare the size of the different matter variables appearing in the PPN formalism, we say that the velocity $v$ is $\mathcal{O}(1)$ and in general $v^n$ is $\mathcal{O}(n)$.

In the PPN approximation \cite{will2014},  the metric can be perturbatively expanded around Minkowski spacetime, i.e.
\bqn
g_{\mu\nu} = \eta_{\mu\nu} + h_{\mu\nu},
\eqn
where $\eta_{\mu\nu} = {\rm diag}(-1,\; 1,\; 1,\; 1)$. In the standard PPN formalism, the metric is required to the PN orders  as \cite{will2014}
\bqn
h_{00} &\sim& \mathcal{O}(2) + \mathcal{O}(4),\\
h_{0i} &\sim & \mathcal{O}(3), \\
h_{ij} &\sim & \mathcal{O}(2).
\eqn
Following the standard PPN formalism, the metric can be related to a series super-potentials,
\bqn
&&h_{00} [\mathcal{O}(2)]: U, \lb{h_SP1}\\
&& h_{00} [\mathcal{O}(4)]: U^2, \Phi_W, \Phi_1, \Phi_2, \Phi_3, \Phi_4, \mathfrak{A}, \mathfrak{B}, \lb{h_SP2}\\
&& h_{0i} [\mathcal{O}(3)]: V_i, W_i, \lb{h_SP3}\\
&& h_{ij} [\mathcal{O}(2)]: U \delta_{ij}, \chi_{, ij}, \lb{h_SP4}
\eqn
where the super-potentials $U$, $\Phi_W$, $\Phi_1$, $\Phi_2$, $\Phi_3$, $\Phi_4$, $\mathfrak{A}$, $\mathfrak{B}$, $V_i$, $W_i$, and $\chi$ are given  by Eqs. (\ref{SPU} - \ref{SPchi}) in Appendix C.

Similar to the previous works \cite{Crisostomi:2017ugk,TZ2019,Qiao:2019hkz}, we consider the unity gauge $\phi=\phi(t)$, i.e., the scalar field $\phi$  is time varying but spatially homogeneous. This scalar field can be either a quintessence field or some other field, which should be determined by the cosmological solution. One of the remarkable features of this scalar field is that it provides a preferred time direction associated with cosmic expansion. This implies this theory is a Lorentz-breaking theory with a parity-violating sector. In this sense, the ghost-free PV gravity does not admit the full diffeomorphism invariance of the four-dimensional spacetime. In other words, the theory only contains the time reparametrization symmetry and the three-dimensional spatial diffeomorphism,
\bqn
\tilde t \to \tilde t &=& t- f(t),\\
\tilde x^i \to \tilde x^i &=& x^i - \xi^i(t, x^i).
\eqn
Under this gauge transformation, we find that the metric transforms as
\bqn
\tilde h_{\tilde{0} \tilde{0}} &=& h_{00} -  2 \lambda_2 (U^2+\Phi_W -\Phi_2) + 2 \dot f,\\
\tilde h_{\tilde{0} \tilde{j}} &=& h_{0j} - \lambda_2 \chi_{,0 j},\\
\tilde h_{\tilde{i} \tilde{j}} &=& h_{ij} - 2 \lambda_2 \chi_{, ij},
\eqn
{where in writing the above expressions, we had chosen $\xi_j = \lambda_2 \chi_{,j}$ \cite{will2014} with $\lambda_2$ being an arbitrary constant. Clearly, by properly choosing $\lambda_2$ we can eliminate the anisotropic term $\chi_{, ij}$ as it was done in the standard post-Newtonian gauge \cite{will2014}.} However, since now $f(t)$ is a function of $t$ only, we cannot eliminate the $\mathfrak{B}$ term in $h_{00}$. Therefore, the general metric coefficients up to $\mathcal{O}(4)$ order in the ghost-free PV theory are given by,
\bqn\lb{ppe}
g_{00} &=& -1 + 2 U - 2 \beta U^2 -2 \xi \Phi_W +\left(2+2 \gamma+\alpha_{3}+\zeta_{1}-2 \xi\right) \Phi_{1} +2\left(1+3 \gamma-2 \beta+\zeta_{2}+\xi\right) \Phi_{2} \nb \\
&&+2\left(1+\zeta_{3}\right) \Phi_{3}+2\left(3 \gamma+3 \zeta_{4}-2 \xi\right) \Phi_{4} -\left(\zeta_{1}-2 \xi\right) \mathfrak{A}+\zeta_{B} \mathfrak{B},\nb \\
g_{0i} &=& -\frac{1}{2}\left(3+4 \gamma+\alpha_{1}-\alpha_{2}+\zeta_{1}-2 \xi\right) V_{i} -\frac{1}{2}\left(1+\alpha_{2}-\zeta_{1}+2 \xi\right) W_{i} +\kappa X_i, \nb \\
g_{i j} &=&(1+2 \gamma U) \delta_{i j},
\eqn
where $(\beta, \gamma, \xi, \zeta_1, \zeta_2, \zeta_3, \zeta_4, \alpha_1, \alpha_2, \alpha_3)$ are 10 PPN parameters in the standard PPN formalism.  Besides these 10 PPN parameters, we also introduce additional contributions to the post-Newtoanian metric, $\zeta_{\rm{B}}$ and $\kappa$. These are different from the standard PPN  formalism in which full diffeomorphism in 4-dimensional spacetime is used to fix the gauge and the parity symmetry is conserved. The introduction of $\zeta_{\rm B}$ is due to the breaking of the Lorentz symmetry of the theory. This is very similar to the case in the Horava-Lifshitz gravity, in which the theory is only spatial covariant and a preferred time direction is chosen \cite{Lin:2013tua}. The term $\kappa$ represents the contributions from the parity violation of the theory, whose explicit form will be determined later in this paper.

Now, we would like to perform a PN expansion of the field equations and obtain a solution in the form of a PN series. From this solution, we are able to read off the PPN parameters by comparing them to the standard PPN super-metric. Before doing so, let us first consider expanding the Ricci, Cotton, $A$, and $B$ tensors to second order. Based on the second-order expansion of these tensors, each component of them can be obtained and the field equation can be solved.
Meanwhile, in order to simplify calculation, we impose the following gauge conditions
\bqn\lb{LMPVu}
 h_{ik}{}^{,k} - \frac{1}{2}h_{,i} &=& \mathcal{O}(4),   \nb\\
  h_{0k}{}^{,k} - \frac{1}{2}h_{,0} &=& \mathcal{O}(5).
\eqn
In the PPN gauge, combined with the above gauge conditions, the components of Ricci tensor and Cotton tensor read \cite{Stephon2007}
\bqn
R_{00} &=&  -\frac{1}{2}\nabla^2h_{00} - \frac{1}{2}h_{00,i}h_{00}{}^{,i} + \frac{1}{2}h^{ij}h_{00,i,j} + \mathcal{O}(6), \nb\\
R_{0i} &=& -\frac{1}{2}\nabla^2h_{0i} - \frac{1}{4}h_{00,0,i} +\mathcal{O}(5), \nb\\
R_{ij}  &=& -\frac{1}{2} \nabla^2 h_{ij} + \mathcal{O}(4); \\
C_{00} &=&\mathcal{O}(6), \nb\\
C_{0i} &=& -\frac{1}{4}\vartheta' \epsilon^{0 kl}{}_{i}\nabla^2 h_{0 l,k} + \mathcal{O}(5),\nb\\
C_{ij} &=& -\frac{1}{2}\vartheta' \epsilon^{0 kl}{}_{(i}\nabla^2 h_{j)l,k} + \mathcal{O}(4);
\eqn
 while the $A_{\mu\nu}$ tensor and $B_{\mu\nu}$ tensor reduce to
\bqn\lb{ABex}
A_{00} &=&\mathcal{O}(6),\nb\\
A_{0i} &=& -\frac{1}{4} \varTheta' \epsilon^{0kl}{}_{i}\nabla^2 h_{0 l,k} + \mathcal{O}(5) ,\nb\\
A_{ij} &=& - \frac{1}{2} \varTheta' \epsilon^{0 kl}{}_{(i}\nabla^2 h_{j)l,k} + \mathcal{O}(4) ;\nb\\
B_{00} &=&\mathcal{O}(6),\nb\\
B_{0i} &=&\mathcal{O}(5) ,\nb\\
B_{ij} &=&  \mathcal{O}(4) ;
\eqn
where $\nabla^2=\eta^{ij} \partial_{i}\partial_{j}$ is the Laplacian of flat space, and
\bqn\lb{Th}
\varTheta=(- 2a_2 + a_3 -8 a_4)\phi'^2.
\eqn
\red{In Appendix A, we have presented the expressions of tensor $A_{\mu\nu}$, $B_{\mu\nu}$ in the form of $A^{(n)}_{\mu\nu}(n=1, 2, 3, 4)$  and $B^{(m)}_{\mu\nu}(m=1, 2, ..., 7)$, where $A^{(n)}_{\mu\nu}$ and $B^{(m)}_{\mu\nu}$ correspond to the components in Lagrangian $\mathcal{L}_{\rm PV1}$ and $\mathcal{L}_{\rm PV2}$, respectively.}
From Eq.(\ref{ABex}) and Eq.(\ref{Th}) we observe that the first term $A^{(1)}_{\mu\nu}$ of $A_{\mu\nu}$ and $B_{\mu\nu}$ tensor have no contributions to the PN expansions or  their contributions are higher than the required order. This implies that the perturbative field equations are modified by the Cotton tensor and the rest terms of $A_{\mu\nu}$ tensor.
The more intuitive explanation for this is because we use the unitary gauge, when the indicators of the scalar field and the Riemann tensor are contracted, a time derivative must appear in the Riemann tensor. It can be seen from Eq.(\ref{LMPV1}) that the two scalars in the first formula are mutually contracted with the two Riemann tensors, that is, there are two-order time derivatives, and only one scalar and Riemann tensor are contracted to each other in the last three formulas, which is naturally lower order than the previous formula.
 A similar analysis is carried out for Eq.(\ref{LMPV2}), taking $M_1$ as an example, where the first derivative of the scalars are contracted with the Riemann tensor and the Levi-Civit\'{a} tensor respectively, so that one of the two tensor indicators must be zero. At the same time, the Riemann tensor and the Levi-Civit\'{a} tensor are contracted with the second derivative of scalar $\phi^{\sigma}_{\nu}$, which will cause a connection $\Gamma^{\sigma}_{\nu 0}$ to appear or the Levi-Civit\'{a} tensor to zero. This combination of connection $\Gamma^{\sigma}_{\nu 0}$ and Riemann tensor is consistent with the lowest order of the first formula in Eq.(\ref{LMPV1}). When the Levi-Civit\'{a} tensor is zero, it is necessary to consider the higher-order approximation in the PPN approximation, such as $M_6$,  the Levi-Civit\'{a} tensor is contracted with two the first derivative of the scalars so that the Levi-Civit\'{a} tensor is zero. In this way, we infer that the lowest order of each term in Eq.(\ref{LMPV2}) is at least the same as or higher than that of the first formula in Eq.(\ref{LMPV1}). This also intuitively explains that their final contribution to the perturbative field equations in the standard first-order PPN approximation does not exist. Therefore, the contribution of the perturbative field equations only comes from tensor $C_{\mu\nu}$, ${A^{(2)}_{\mu\nu}}$, ${A^{(3)}_{\mu\nu}}$ and ${A^{(4)}_{\mu\nu}}$.

The stress-energy tensor $T^m_{\mu\nu}$ of the matter field is given by
\bqn
T^m_{00} &=&\rho \left( 1 + \Pi + v^2 - h_{00}\right) + \mathcal{O}(6),\nb\\
T^m_{0i} &=& -\rho v_i + \mathcal{O}(5) ,\nb\\
T^m_{ij} &=& \rho v_i v_j + p\delta_{ij} + \mathcal{O}(6) .
\eqn
What we need for the subsequent calculations of the metric equation is the expansion of the trace-reversed energy-momentum tensor, which is expressed as
\bqn
S^m_{00} &=& \frac{1}{2}\rho + \frac{1}{2}\rho\Pi + \rho v^2 + \frac{3}{2} p - \frac{1}{2}\rho h_{00} + \mathcal{O}(6),\nb\\
S^m_{0i} &=&- \rho v_i + \mathcal{O}(5) ,\nb\\
S^m_{ij} &=& \frac{1}{2}\rho \delta_{ij} + \rho v_i v_j + \frac{1}{2}\rho\Pi\delta_{ij} + \frac{1}{2}\rho h_{ij} - \frac{1}{2} p\delta_{ij} + \mathcal{O}(6) .
\eqn
We also know the tensor $T^{\phi}_{\mu\nu}$ of the scalar field by Eq.(\ref{sft}), which is expanded in the form
\bqn
T^{\phi}_{00} &=&\frac{1}{2} \phi'^2 \left( 1 - 2 h_{00} \right) + (-1 + h_{00})V(\phi) + \mathcal{O}(6),\nb\\
T^{\phi}_{0i} &=& -\frac{1}{2} \phi'^2 h_{0i} + V({\phi} ) h_{0i} + \mathcal{O}(5) ,\nb\\
T^{\phi}_{ij} &=& \frac{1}{2} (-\delta_{ij} - h_{ij} + h_{00}\delta_{ij} + h_{00}h_{ij} + h_{00}h_{00}\delta_{ij} ) \phi'^2 + V(\phi)\delta_{ij} + V(\phi)h_{ij} + \mathcal{O}(6) ,
\eqn
so we can obtain
\bqn\lb{sfe}
S^{\phi}_{00} &=&V(\phi) - h_{00}V(\phi) + \mathcal{O}(6),\nb\\
S^{\phi}_{0i} &=&-h_{0i}V(\phi) + \mathcal{O}(5) ,\nb\\
S^{\phi}_{ij} &=& -\delta_{ij}V(\phi) - h_{ij}V(\phi)+ \mathcal{O}(4) .
\eqn
Now, we have the expanded components of all tensors contained in the metric field equation (\ref{feq}). By substituting these expansions into the metric field equations, we can present a similar expansion of the metric field equations, and decompose them into different velocity orders. In the following section, we shall study these equations in detail in different orders.

\section{Parametrized post-Newtonian solution \label{sec4}}
\renewcommand{\theequation}{4.\arabic{equation}} \setcounter{equation}{0}

\subsection{ Zeroth order metric and scalar equations}

To the zeroth-order, from metric field equation (\ref{feq}) we obtain
\bqn
\eta_{\mu\nu}V(\phi)=0,  \Rightarrow V(\phi)=0.
\eqn
Combining the Eq.(\ref{sfe}), this solution indicates that each expanded component of the tensor $S^{\phi}_{\mu\nu}$ does not contribute to the metric equation.

\subsection{ $h_{00}$ and $h_{ij}$ to $\mathcal{O}(2)$}

At  the second velocity order, we obtain the $00$-metric equation
\bqn
-\frac{1}{2}\nabla^2 h_{00} = 4\pi G\rho,
\eqn
this is a Possion equation whose solution is
\bqn\lb{eq0}
h_{00} = 2U + \mathcal{O}(4).
\eqn
The $ij$-equations to same order  is given by
\bqn\lb{ijeq}
-\frac{1}{2} \nabla^2 h_{ij} -\frac{1}{2} f' \epsilon^{0 kl}{}_{(i}\nabla^2 h_{j)l,k}  =  4\pi G \rho \delta_{ij},
\eqn
where
\bqn\lb{f.para}
f=\vartheta + \Theta = \vartheta + (- 2a_2 + a_3 -8 a_4)\phi'^2
\eqn
is only time-dependent. Note that, as mentioned above, the second tensor contribution on the left hand comes from the tensor $C_{\mu\nu}$, ${A^{(2)}_{\mu\nu}}$, ${A^{(3)}_{\mu\nu}}$ and ${A^{(4)}_{\mu\nu}}$. It is clear that Eq.(\ref{f.para}) just contains the coupling parameters $\vartheta$, $a_2$, $a_3$ and $a_4$, corresponding to the contributions of the tensor $C_{\mu\nu}$, ${A^{(2)}_{\mu\nu}}$, ${A^{(3)}_{\mu\nu}}$ and ${A^{(4)}_{\mu\nu}}$, respectively.
In order to solve the above equation, we can do similar processing according to Ref. \cite{Stephon2007}. We introduce a effective metric $\mathcal{H}_{ij}$ to rewrite Eq.(\ref{ijeq}) as
\bqn\lb{eeq}
\nabla^2 \mathcal{H}_{ij} = -8 \pi G \rho \delta_{ij},
\eqn
where $\mathcal{H}_{ij} =  h_{ij} + f' \epsilon^{0 kl}{}_{(i} h_{j)l,k}$. We can easily get the solution
\bqn\lb{eme}
\mathcal{H}_{ij} = 2U \delta_{ij}+ \mathcal{O}(4).
\eqn
With the form of effective metric, then we need to obtain the actual metric. Combining the Eq.(\ref{eeq}) with Eq.(\ref{eme}), we have the following differential equation
\bqn\lb{eq1}
h_{ij} +  f' \epsilon^{0 kl}{}_{(i} h_{j)l,k} = 2U \delta_{ij}.
\eqn
Obviously, the above equation has a solution whose zeroth-order term is that predicted by GR and the PV term is a perturbative correction. So we make the ansatz
\bqn
h_{ij} = 2U \delta_{ij}+ f'\zeta_{ij},
\eqn
where we assume that $\zeta \sim \mathcal{O}(f')^0$. Substituting this ansatz to Eq.(\ref{eq1}) we get
\bqn
\zeta_{ij} + f' \epsilon^{0 kl}{}_{(i} \zeta_{j)l,k} =0.
\eqn
Note that, the second term on the left-hand side generates a second-order correction. Ignoring the second-order term, one can find that $\zeta_{ij}$ vanishes to this order.
Thus, the special metric perturbation to $\mathcal{O}(2)$ is directly given by the GR prediction without any parity violation correction, namely
\bqn\lb{eq2}
h_{ij} = 2U\delta_{ij} + \mathcal{O}(4).
\eqn

\subsection{ $h_{0i}$ to $\mathcal{O}(3)$}

The $0i$-equation to $\mathcal{O}(3)$ is given by
\bqn
-\frac{1}{2}\nabla^2h_{0i} - \frac{1}{4}h_{00,0,i} - \frac{1}{4} f' \epsilon^{0 kl}{}_{i}\nabla^2 h_{0 l,k} = -8\pi G \rho v_i.
\eqn
Using the above solution of $h_{00}$ to $\mathcal{O}(2)$ and the effective metric we have
\bqn
\nabla^2 \mathcal{H}_{0i} + U_{,0,i} = 16 \pi G \rho v_i,
\eqn
where the form of this effective metric is $\mathcal{H}_{0i} =  h_{0i} + \frac{1}{2} f' \epsilon^{0 kl}{}_{i} h_{0l,k}$. Comparing this equation with standard GR field equation to $\mathcal{O}(3)$,  the solution is given by
\bqn
\mathcal{H}_{0i} = -\frac{7}{2} V_i - \frac{1}{2}W_i.
\eqn
Inserting this solution into the form of the effective metric, we have
\bqn\lb{oieq}
h_{0i} + \frac{1}{2} f' \epsilon^{0 kl}{}_{i} h_{0l,k} =  -\frac{7}{2} V_i - \frac{1}{2}W_i.
\eqn
Likewise, the solution is consist of the GR prediction and perturbative correction, namely
\bqn
h_{0i}  =  -\frac{7}{2} V_i - \frac{1}{2}W_i + f' \zeta_i,
\eqn
where we assume that $\zeta$ is of $\mathcal{O}(f')^0$. Eq.(\ref{oieq}) reduces to
\bqn
\zeta_i + \frac{1}{2}f' (\nabla \times \zeta)_i = \frac{1}{2} \left[ \frac{7}{2} (\nabla \times V)_i + \frac{1}{2} (\nabla \times W)_i \right],
\eqn
where $(\nabla \times \zeta)_i = \epsilon^{ kl}{}_{i} \zeta_{l,k}$. \red{Note that, we may neglect the second term on the left-hand side,} which is also a second-order correction.  Meanwhile, since the curl of the $V_i$ potential is equal to the curl of the $W_i$ potential, we can obtain the actual perturbative metric
\bqn\lb{eq3}
h_{0i} = -\frac{7}{2} V_i - \frac{1}{2}W_i + 2f' (\nabla \times V)_i +\mathcal{O}(5).
\eqn

\subsection{ $h_{00}$  to $\mathcal{O}(4)$}

The fourth-order metric field equation reads
\bqn
 -\frac{1}{2}\nabla^2h_{00} - \frac{1}{2}h_{00,i}h_{00}{}^{,i} + \frac{1}{2}h^{ij}h_{00,i,j} =  8\pi G \left( \frac{1}{2}\rho\Pi + \rho v^2 + \frac{3}{2} p - \frac{1}{2}\rho h_{00} \right),
\eqn
where the 00-component of the field equation to $\mathcal{O}(4)$ is only GR prediction without any other correction. The solution of this modified gravity is same to GR, which is given by
\bqn\lb{eq4}
h_{00} = -2U^2 + 4\Phi_1 + 4\Phi_2 + 2\Phi_3 + 6\Phi_4 + \mathcal{O}(6).
\eqn

Having obtain all the necessary perturbative solutions of the metric field equations as given by Eq.(\ref{eq0}), Eq.(\ref{eq2}), Eq.(\ref{eq3}) and Eq.(\ref{eq4}), let us write the full metric of this modified gravity:
\bqn\lb{ppe1}
g_{00} &=& -1+ 2U -2U^2 + 4\Phi_1 + 4\Phi_2 + 2\Phi_3 + 6\Phi_4 + \mathcal{O}(6), \nb\\
g_{0i} &=& -\frac{7}{2} V_i - \frac{1}{2}W_i + 2f'(\nabla \times V)_i +\mathcal{O}(5), \nb\\
g_{ij} &=& (1+2U) \delta_{ij} + \mathcal{O}(4).
\eqn
where $f  = \vartheta + (- 2a_2 + a_3 -8 a_4)\phi'^2$.

We can read off the PPN parameters of this model by comparing Eq.(\ref{ppe}) to Eq.(\ref{ppe1}), the values of the PPN parameters are given as following
\bqn
\gamma &=& \beta =1,\nb\\
\xi &=& \alpha_1 = \alpha_2 = \alpha_3 = \zeta_1 = \zeta_2 = \zeta_3 = \zeta_4 = \zeta_B=0,\nb\\
\kappa&=&2f'.
\eqn
It can be observed that all the 11 PPN parameters have the same values as GR. However, Eq.(\ref{ppe1}) contains an extra term $\kappa X_i \equiv2f'(\nabla \times V)_i$ that cannot be modeled by the standard PPN metric of Eq.(\ref{ppe}), namely a \red{curl-type} term contribute to $g_{0i}$. Therefore, the parity-violating terms of this ghost-free PV gravity only contribute to $0i$-components of the PPN metric in the PPN approximation. As expected, when $a_i=0$, this \red{curl-type} term reduces to the same form as that in Chern-Simons gravity \cite{Stephon2007a,Stephon2007}.

\section{Precession of Orbiting Gyroscopes\label{sec5}}
\renewcommand{\theequation}{5.\arabic{equation}} \setcounter{equation}{0}


In previous discussions, we find that in comparison with GR, the modification of the ghost-free PV gravity only affects the gravitomagnetic sector of the metric in the PPN approximation. \red{In GR, the gravitomagnetic sector of the spacetime arises from the rotation of the gravitational object and results in a frame-dragging effect on the objects orbiting it. For example, for a gyroscope orbiting the Earth, the frame-dragging effects due to the rotation of the Earth can produce an extra procession of the gyroscope spin axis.} Obviously, the relevant experiment of the frame-dragging effect \red{by using gyroscopes in the near-earth artificial satellites} can be used to test \red{or constrain the gravitomagnetic effects due to the parity violation in} this kind of gravitational theory.  \red{One of the important experiments for testing the frame-dragging effect is performed by the GPB experiment \cite{GPB1}. The experiment designed four gyroscopes to be mounted on the GPB satellite, which was in a polar orbit 640 km from the Earth's surface and had an orbital time period of 97.65 min. As the satellite orbits, the spin direction of the gyroscopes changes due to the geodesic and frame-dragging effect. Both effects \red{have been detected} by measuring changes in the direction of the gyroscope's spin axis, which are in agreement with the predictions of GR.} To achieve the constraint of the new correction  \red{from parity violation with the measurement of frame-dragging effect by GPB}, we investigate the spin precession of a gyroscope loaded on the satellite which orbits a weakly-graviting and slowly-rotating object.

 {Considering an orbiting, rotating, nearly spherical gravitational object in the standard PPN point-particle approximation, and here we take the Earth as an example, the PPN vector potentials can be written as} \cite{will2018th}
\bqn
V^i &=& \frac{M_{\oplus}}{r} v_{\oplus}^i + \frac{1}{2} \left( \frac{J_{\oplus}}{r^3} \times x \right)^i,
\eqn
{where $M_{\oplus} $, $v_{\oplus}$, $J_{\oplus}^i$ are the mass, the orbital velocity, the spin-angular momentum of the Earth respectively}, $x^i$ and $|x^i| = r$ denote the satellite-Earth vector and distance. Using the vector potential $V_i$, we can calculate the specific expression of the metric correction term.
In order to facilitate understanding and distinction, we redefine the correction of the metric to GR as $\delta g_{0i} \equiv g_{0i} - g^{\rm{GR}}_{0i} = \kappa X_i$, and $g^{\rm{GR}}_{0i}$ is the standard GR metric.  Substituting the above vector potential $V_i$ to the correction $\delta g_{0i}$, we can obtain
\bqn
\delta g_{0i} = 2\frac{f'}{r} \left[ \frac{M_{\oplus}}{r} \left( v_{\oplus} \times n \right)^i - \frac{J^i_{\oplus}}{2r^2} + \frac{3}{2}\frac{\left(  J_{\oplus} \cdot n \right)}{r^2}  n^i \right],
\eqn
where $n^i=x^i/r$ is a unit vector, the $\cdot$ and $\times$ operators are the flat-space inner and cross products.

For a free gyroscope with a spin three-vector $ \boldsymbol{S} $ in the presence of the gravitational field, the precession of a spin three-vector $\ \boldsymbol{S} $ is expressed as
\bqn
\frac{d \boldsymbol{S} }{dt} =  \boldsymbol{\Omega} \times  \boldsymbol{S} ,
\eqn
where $ \boldsymbol{\Omega} \equiv -\frac{1}{2} \nabla \times  \boldsymbol{g} $ is the precession rate and $ \boldsymbol{g}=g_{0i}$.
{Therefore, the precession rate caused by the term of $g_{0i}$ is given by $\boldsymbol{\Omega}=\boldsymbol{\Omega}^{\rm GR}+\delta\boldsymbol{\Omega}$, where $\boldsymbol{\Omega}^{\rm GR}$ and $\delta\boldsymbol{\Omega}$ denote the contributions in GR and the modification respectively. The expressions of two terms are given by
\bqn
\boldsymbol{\Omega}^{\rm GR} &\equiv& { \frac{1}{r^3}\left[ 3(\boldsymbol{J}_{\oplus} \cdot \boldsymbol{n})\boldsymbol{n} - \boldsymbol{J}_{\oplus} \right] },\\
 \delta  \boldsymbol{\Omega} &\equiv & -\frac{1}{2} \nabla \times \delta \boldsymbol{g}
 = \frac{f' M_{\oplus}}{r^3} \left[ \boldsymbol{v}_{\oplus} -   3\left(  \boldsymbol{v}_{\oplus} \cdot  \boldsymbol{n} \right) \boldsymbol{n} \right] \lb{mpr},
\eqn
where $\delta  \boldsymbol{g} \equiv \delta g_{0i}$. {Note that, this is a significant difference between GR and the ghost-free PV gravity. GR's prediction is associated with spin $\boldsymbol{J}_{\oplus}$ of the Earth, which means that the term $g_{0i}$ in GR contributes only the frame-dragging precession. While in the ghost-free PV gravity, the term $g_{0i}$ contributes both the frame-dragging precession and the geodetic precession. The former is exactly same with that in GR, since the effects of spin $\boldsymbol{J}_{\oplus}$ to the modification of precession rate cancel each other out in Eq.(\ref{mpr}). The latter is the modification caused by PV term in this theory, which is related to the mass $M_{\oplus}$ and orbital velocity $\boldsymbol{v}_{\oplus}$ of the Earth.
}}

{In the classical GR's prediction \cite{EP.Will2014}, in order to maximize the geodesic precession, it is necessary to select the spin $\boldsymbol{S}$ on the orbital plane. Meanwhile, in order to distinguish the spin-spin precession and the spin-orbit precession, the optimal satellite orbit choice is a polar orbit (that is, the tilt angle of the satellite orbit is $\iota=\pi/2$), which is adopted in the GPB experiment.}  Similar to previous work \cite{EP.Will2014}, we use a Cartesian coordinate system $( \boldsymbol{e}_p, \boldsymbol{e}_q, \boldsymbol{e}_z)$ to describe the gyroscope's orbital plane, where $\boldsymbol{e}_z$ is the unit normal direction of the plane, $\boldsymbol{e}_p$ unit vector lies in the plane and points toward the ascending node, and $\boldsymbol{e}_q$ unit vector is in the same direction as the spin $\boldsymbol{J}_{\oplus}$ of the Earth and orthogonal to the first two.
{The direction of correction is determined by the spin three-vector $\boldsymbol{S}$ of the gyroscope and the orbital velocity vector $\boldsymbol{v}_{\oplus}$ of the Earth, so we need to define a parameter $\gamma$, which denotes the angle between the polar orbital plane of the satellite and the direction of the Earth's translational (or orbital) speed $\boldsymbol{v}_{\oplus}$.
Meanwhile,  since the Earth's axial has an angle of inclination on the ecliptic surface, there is an angle $\theta=66.5^{\circ}$ between the spin $\boldsymbol{J}_{\oplus}(\boldsymbol{e}_q)$ and velocity vectors $\boldsymbol{v}_{\oplus}$ of the Earth.}
 In this satellite orbital coordinate system, there are expressions for correction:  $\boldsymbol{n}={ \rm cos} \mathcal{F} \boldsymbol{e}_p + {\rm sin} \mathcal{F} \boldsymbol{e}_q$; $\boldsymbol{v}_{\oplus}=v_{\oplus}({\sin\theta\cos}\gamma \boldsymbol{e}_p +\cos\theta\boldsymbol{e}_q+ \sin\theta\sin\gamma \boldsymbol{e}_z)$, in which $\mathcal{F} = 2 \pi  t/P$ is the satellite orbital phase, $P = 2 \pi r^{3/2} (GM_{\oplus})^{-1/2} $ is the orbital period.
Substituting the foregoing relations into the modification precession rate (\ref{mpr}), we obtain
\bqn\lb{Omega5}
\delta  \boldsymbol{\Omega}
 &=& \frac{f' M_{\oplus} v_{\oplus}}{r^3} \Bigg[ -\frac{1}{2}( \sin\theta\cos\gamma+3{\cos}2 \mathcal{F}\sin\theta\sin\gamma+\sin 2 \mathcal{F} \cos\theta) \boldsymbol{e}_p \nb\\
  &&-\frac{1}{2} (\cos\theta+3\sin2\mathcal{F}\sin\theta\sin\gamma-3\cos 2\mathcal{F}\cos\theta )\boldsymbol{e}_q + \sin\theta \sin \gamma \boldsymbol{e}_z \Bigg].
\eqn
Since this precession is very small, the initial value $\boldsymbol{S}_0$ of the gyroscope spin can be used instead of $\boldsymbol{S}$. We only focus on the cumulative effect, which is the non-oscillating terms in the formula (\ref{Omega5}). Taking the average of the precession equation for a complete circular orbital motion period, we can get
\bqn\lb{do5}
\left < \delta\frac{d\boldsymbol{S}}{dt} \right>=\frac{f' M_{\oplus} v_{\oplus}}{r^3} \left[ -\frac{1}{2} {\sin\theta\cos}\gamma\boldsymbol{e}_p-\frac{1}{2} \cos\theta \boldsymbol{e}_q + \sin\theta\sin\gamma \boldsymbol{e}_z \right] \times \boldsymbol{S}_0.
\eqn
Since the spin of the gyroscope lies in the orbital plane, we have
\bqn\lb{s0}
\boldsymbol{S}_0=\cos\psi \boldsymbol{e}_p + \sin \psi \boldsymbol{e}_q,
\eqn
where $\psi$ is the angle between $\boldsymbol{S}_0$ and the line of node. Combining the formulas (\ref{do5}) and (\ref{s0}), we have
\bqn
 \left < \delta\frac{d\boldsymbol{S}}{dt} \right> &=& \frac{f' M_{\oplus} v_{\oplus}}{r^3} \left[ -\sin \theta \sin \gamma \sin \psi \boldsymbol{e}_p + \sin\theta\sin \gamma \cos \psi \boldsymbol{e}_q +\frac{1}{2} (\cos\theta\cos\psi-\sin\theta\cos \gamma\sin\psi ) \boldsymbol{e}_z \right]  \nb\\
 &\equiv&  \left < \delta\frac{d\boldsymbol{S}}{dt} \right>_{\rm NS}+\left < \delta\frac{d\boldsymbol{S}}{dt} \right>_{\rm WE},
\eqn
where
\bqn
\left < \delta\frac{d\boldsymbol{S}}{dt} \right>_{\rm NS}&=&\frac{f' M_{\oplus} v_{\oplus}}{r^3} \left( -\sin \theta \sin \gamma \sin \psi \boldsymbol{e}_p + \sin\theta\sin \gamma \cos \psi \boldsymbol{e}_q \right) \lb{NS},\\
\left < \delta\frac{d\boldsymbol{S}}{dt} \right>_{\rm WE}&=&\frac{f' M_{\oplus} v_{\oplus}}{2r^3}(\cos\theta\cos\psi-\sin\theta\cos \gamma\sin\psi ) \boldsymbol{e}_z \lb{WE}.
\eqn
{Eq.(\ref{NS}) and Eq.(\ref{WE}) represent the corrections in the same direction as the geodetic effect and the Lense-Thirring effect respectively, which are also known as the North-South(NS) and West-East (WE) direction in GPB experiment. Thus, we can obtain the rate of the correction angular precession in two directions are given by}
\bqn
\left|\left < \delta\frac{d\boldsymbol{S}}{dt} \right>\right|_{\rm NS} &=&  \frac{f' M_{\oplus} v_{\oplus}}{r^3} |\sin\theta \sin \gamma | \lb{5.12} , \\
\left|\left < \delta\frac{d\boldsymbol{S}}{dt} \right>\right|_{\rm WE} &=&  \frac{f' M_{\oplus} v_{\oplus}}{2r^3} |\cos\theta\cos\psi-\sin\theta\sin\psi \cos \gamma | \lb{5.13}.
\eqn
From the above two results, we can see that except for the parameter $\gamma$ which is not determined, the other parameters are all known constants. In order to analyze a more specifically, we decompose the vector $\boldsymbol{v}_{\oplus}$ into vector $\boldsymbol{v}_{T}= v_{\oplus}\sin\theta({\cos}\gamma \boldsymbol{e}_p+ \sin\gamma \boldsymbol{e}_z)$ and vector $\boldsymbol{v}_{S}= v_{\oplus}\cos\theta\boldsymbol{e}_q$. The vector $\boldsymbol{v}_{S}$ is a constant vector  with a fixed direction. The magnitude and direction of vector $\boldsymbol{v}_{T}$ change with the movement of the Earth, resulting in a cyclical change in the angle $\gamma$ that the period is the period of the Earth's orbit. We define $\delta\gamma$ as the change of the angle $\gamma$ with the movement of the Earth, that is, $\gamma=\gamma_0+\delta\gamma$, $\delta\gamma=2\pi t/T$ and  $T=1~{\rm yr}$. We have $\delta\gamma=2\pi P/T \simeq 0.068^{\circ}$  in a complete circular satellite orbital motion period of the GPB experiment, which is a negligible variation in Eq.(\ref{5.12}) and Eq.(\ref{5.13}).
\red{However, the GPB experimental results are given in terms of mas/yr. We need to integrate Eq.(\ref{5.12}) and Eq.(\ref{5.13}) over a year and then average. This time interval is exactly the change period of the parameter $\gamma$, and the items including $\sin\gamma$ and $\cos\gamma$ are both zero. Therefore, in the case of complete Earth's orbital cycle, the terms containing $\gamma$ are oscillating terms, whose contributions can be ignored in Eq.(5.12) and Eq.(5.13).} Thus, within a complete Earth's orbital cycle, the rate of the correction angular precession in only one direction is given by
\bqn
\left|\left < \delta\frac{d\boldsymbol{S}}{dt} \right>\right|_{\rm WE} &=&  \frac{f' M_{\oplus} v_{\oplus}}{2r^3} \cos\theta\cos\psi .
\eqn


{GPB experiment has successfully tested the rate of the angular precession of GR's prediction and the GPB team announced the observation results that the geodetic drift rate is $-6601.8 \pm 18.3 ~{\rm mas/yr}$ and the frame-dragging drift rate is $-37.2 \pm 7.2 ~{\rm mas/yr}$ \cite{GPB1}. The GPB gyroscope is at a radius of $r \sim 7000~ \rm{km} $, $\psi=16.8^{\circ} $ \cite{will2018th}. Therefore, combined with the experimental parameters and results, we can estimate the bound of parameter in orders of magnitude. } The corresponding constraint is given as
\bqn
\left|\left < \delta\frac{d\boldsymbol{S}}{dt} \right>\right|_{\rm WE}   \lesssim  7.2 ~ {\rm mas/yr} ~~~\Rightarrow ~~~ f' \lesssim 10^4 ~{\rm m},
\eqn
{and its associated $energy~ scale$ of parity violation in gravity is $M_{\rm PV}:= 1/f' \gtrsim 10^{-11} {\rm eV}$.} Note that, by GPB observations, the constraint on non-dynamical CS modified gravity is $ M_{\rm CS}:= 1/\vartheta' \gtrsim 10^{-13} {\rm eV}$ \cite{Stephon2007a,Stephon2007,Smith:2007jm}, \red{which differs by two order of magnitude} than our result. This difference is caused by the different accuracy \footnote{At that time, the results of the GPB have yet been announced.} and the inappropriate rough approximation \footnote{The authors assumed the Newtonian limit $ \mathcal{O}(J_{\oplus}) \sim \mathcal{O}(M_{\oplus} v_{\oplus}r)$ in previous works. However, the fact is $ \mathcal{O}( M_{\oplus} v_{\oplus}r/J_{\oplus}) \sim 10^{2}$.} in calculations \cite{Stephon2007a,Stephon2007}.
However, in Ref.\cite{Smith:2007jm}, this is because the authors only consider the stationary, spinning source, that is, do not consider the Earth's translational (or orbital) velocity. The exact solution of the field equation is given precisely, and it is shown that the boundary terms of the spin-related oscillation will also modify the precession. These boundary terms correspond to the higher-order terms of the parameter $f'$, which shows that they are consistent with the assumption we used when solving the field equation Eq.(\ref{oieq}). A more detailed description of this difference is discussed in the next section.
Taking into account these factors, we find the constraint of the ghost-free PV gravity we obtain is consistent with those in previous works.


\section{Conclusions and Discussions \label{sec6}}
\renewcommand{\theequation}{6.\arabic{equation}} \setcounter{equation}{0}

In these discussions, we are mainly to explain that the final precession effect has nothing to do with spin in the ghost-free parity-violating gravity we studied, but in Ref.\cite{Smith:2007jm} the main correction is spin-related.  There are two main reasons for the difference. On the one hand, it is only considering the static and rotating source, ignoring the influence of the source's translation (or orbital motion). There are no translational velocity-related terms in the solution of the corresponding equation. On the other hand, it is considered that the vector potential is continuous inside and outside the source, so the boundary terms are necessary for these two solutions as Eq.(B12) in Ref.\cite{Smith:2007jm}. Even though these reasons have been explained in this reference, we repeat them here to analyze the problem of parameter constraint range.

In  Ref.\cite{Smith:2007jm}, the gravito-electro-magnetic analogy is used to study the problem of the static rotation source. In the Lorenz gauge, the usual gravitomagnetic vector potential is defined, which is related to the metric. At this time, the modified field equation is equivalent to Ampere’s law for Chern-Simons gravity. The gravitomagnetic vector potential is given as Eq.(B12) in Ref.\cite{Smith:2007jm}. Compared with Ref.\cite{Stephon2007a,Stephon2007}, this gravitomagnetic vector potential does not contain the terms of translational velocity, but it contains some oscillating boundary terms.

The modified field equation of the ghost-free PV gravity is consistent with CS gravity, so the difference stems from the solution of Eq.(\ref{oieq}). We have solved the equation according to the method of Ref.\cite{Stephon2007a,Stephon2007}, assuming that the high-order terms related to the parameter $f'$ of the solution are ignored. At the same time, we consider a source of rotation with translational (orbital) velocity. Through analysis, we find that the boundary terms of the gravitomagnetic vector potential in Ref.\cite{Smith:2007jm} are higher-order terms about the parameters $f'$, which shows that the assumption we used is self-consistent.

However, we can get a new gravitomagnetic vector potential by combining the terms of translational velocity with the gravitomagnetic vector potential, and find that the new gravitomagnetic vector potential is still a solution of the equation. Although this solution is still not a complete solution, we can roughly analyze the influence of the oscillating boundary terms and the motion velocity terms on the precession effect at the same time.

\begin{figure*}
  \centering
  \includegraphics[width=8cm]{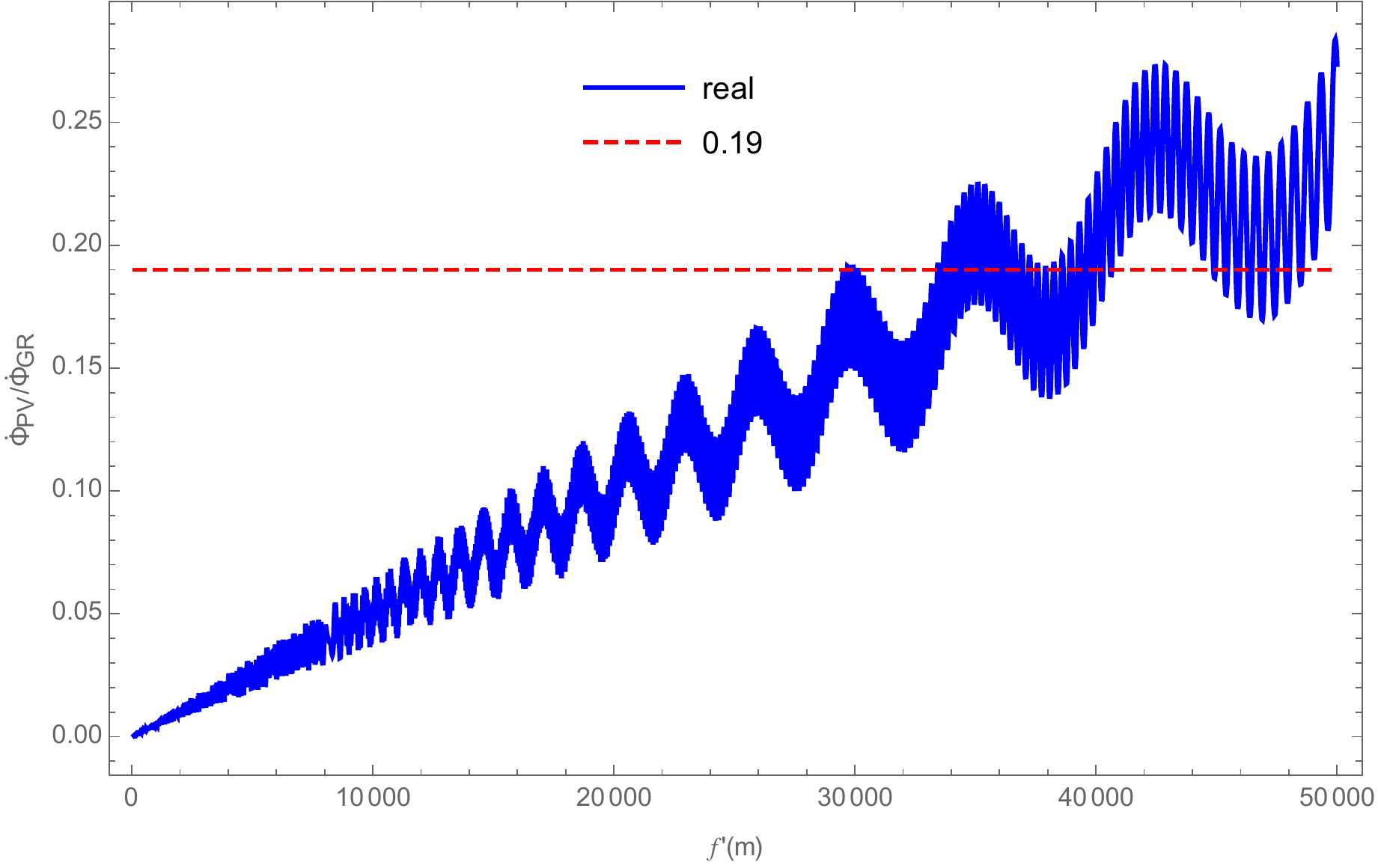}
  \caption{The oscillating blue solid line represents the change of the ratio $\dot{\Phi}_{\rm PV}/\dot{\Phi}_{\rm GR}$ with the parameter $f'$. A 19\% verification of general relativity (the red dotted line) leads to a limit on the parameter $f' \lesssim 10^4 ~{\rm m} $ in GPB.} \label{solution}
\end{figure*}

According to the method of Ref.\cite{Smith:2007jm}, relative to the general relativity result, we find
\bqn\lb{eqd1}
\dot{\Phi}_{\rm PV}/\dot{\Phi}_{\rm GR}=5f'v_{\oplus}\cos\theta/(4R^2\omega_{\oplus}) +15(r^2/R^2)j_2(2R/f')\left[ y_1(2r/f') + (2r/f') y_0(2r/f')\right]
\eqn
where $\dot{\Phi}_{\rm GR}$ represents the precession rate predicted by GR, and $\dot{\Phi}_{\rm PV}$  represents the modified precession rate generated by the PV terms. In Fig.\ref{solution}, we plot Eq. (\ref{eqd1}) for a GPB detection of the gravitomagnetic precession to within 19\% of its value in general relativity. From this figure, we can get a constraint: $f' \lesssim 10^4 ~{\rm m} $, which shows that it is in the same order of magnitude as the result that we do not consider the boundary terms.

In this paper, we investigate the slow-motion and weak-field approximation of the general ghost-free PV theory of gravity in the PPN framework. We derive the perturbative field equations, which use the PPN approximation to expand all tensors in the modified field equations to the second order. By solving the perturbative field equations, we obtain the complete PPN metric and PPN parameters of this theory. Similar to the CS modified gravity, we find that the ghost-free PV theory of gravity produces a new \red{curl-type} term to the gravitomagnetic sector of the metric.  This extra term is the same as in CS gravity except for the coupling parameters. We calculate the modification of the frame-dragging effect, which is caused by the extra term, to the precession of gyroscopes. {We find that this modification contributes in only one direction of the geodetic precession within a complete Earth's orbital cycle, which exactly corresponds to the changes of GR's frame-dragging precession. Using the precision of precession effects from the GPB experiment, we obtain the constraints of the coupling parameters.}

{\color{black}At the end of this paper, we should mention that in addition to the modification on precession effects of gyroscopes, the PV terms in this theory also lead to the modification on the orbital procession of the point particle orbiting a rotating gravitational field \cite{lageos}. Therefore, the measurement of LAGEOS satellites can also follow a constraint on the model parameters. Since this potential constraint is much weaker than that derived from GPB experiment \cite{Smith:2007jm}, we do not consider this effect in this article. Another interesting testbed for this theory is the binary pulsar systems, the PV terms might induce several effects, including the modification on the rate of periastron precession, the modification on the Schiff procession and/or the frame-dragging procession, as well as the modification on Lense-Thirring procession. We leave the detailed study on this topic as future work.}

\section* {Appendix A: The expressions of $A_{\mu\nu}$, $B_{\mu\nu}$ and $F_{,\phi}$\label{Ap1 } }
\renewcommand{\theequation}{A.\arabic{equation}} \setcounter{equation}{0}

In this Appendix, we list the explicit expression of $A_{\mu\nu}$, $B_{\mu\nu}$ and $F_{,\phi}$ in Eqs.(\ref{2.8}) and (\ref{2.12}). The expression of $A_{\mu\nu}$ in Eq.(\ref{2.8}) contains four terms,
\bqn
A_{\mu\nu} = \sum_{A=1}^{4} A_{\mu\nu}^{(A)},
\eqn
where
\bqn
A^{(1)}_{\mu\nu}&=& \varepsilon^{\kappa\gamma\alpha \beta} R_{\alpha \beta \rho \sigma} R_{\kappa\gamma}{}^{\rho}{}_{ \lambda} \phi^\sigma \phi^\lambda a_{1,X}\phi_{(\nu}\phi_{\mu)}
-  a_1 \varepsilon^{\lambda\gamma\alpha\beta} \phi_{\sigma}\phi_{(\mu} R^{\sigma}{}_{\rho\alpha\beta}R^{\rho}{}_{\nu)\lambda\gamma}     \nb\\
&& + 2 \varepsilon^{\sigma\beta\alpha}{}_{(\nu} \nabla_{\rho} \nabla_{\alpha} \left(a_1 \phi_{\mu)}\phi_{\lambda} \right)   R_{\sigma\beta}{}^{\rho\lambda}
   + 2 \varepsilon^{\sigma\beta\alpha}{}_{(\nu} \nabla_{\alpha} \left(a_1 \phi_{\mu)}\phi_{\lambda} \right) \nabla_{\rho} R_{\sigma\beta}{}^{\rho\lambda}    \nb\\
&& + 2 \varepsilon^{\sigma\beta\alpha}{}_{(\nu} \nabla_{\rho} \nabla_{\alpha} \left( a_1\phi_{\lambda}\phi^{\rho}  \right) R^{\lambda}{}_{\mu)\sigma\beta}
   + 2 \varepsilon^{\sigma\beta\alpha}{}_{(\nu} \nabla_{\alpha} \left( a_1\phi_{\lambda}\phi^{\rho} \right) \nabla_{\rho}  R^{\lambda}{}_{\mu)\sigma\beta}    ,\nb\\
   A_{\mu\nu}^{(2)} &=&  -\varepsilon^{\delta\tau\alpha \beta} a_{2,X} \phi_{(\mu} \phi_{\nu)} \phi_\tau \phi_\kappa R^{\sigma}{}_{\rho \alpha \beta } R^{\rho }{}_{\sigma \delta \lambda } g^{\lambda\kappa}
       -\varepsilon^{\lambda\kappa\alpha \beta} \phi_\kappa \phi_{(\nu} a_2 R^{\sigma}{}_{\rho \alpha \beta } R^{\rho }{}_{\sigma \lambda \mu) }   \nb\\
&&  +2\varepsilon^{\sigma \beta\alpha}{}_{ (\nu}  g^{\lambda\kappa}  \nabla_\rho \nabla_\alpha(\phi_\beta \phi_\kappa a_2 R^{\rho }{}_{\mu)\sigma \lambda })
      +\varepsilon^{\rho\lambda\alpha \beta}  \nabla_\sigma \nabla_\rho (\phi_\lambda \phi_{(\nu} a_2 R^{\sigma}{}_{\mu)\alpha \beta })   \nb\\
&&  -\varepsilon_{(\mu}{}^{\rho\alpha \beta}  g^{\lambda\kappa}  \nabla_\sigma \nabla_\lambda (\phi_\rho \phi_\kappa a_2 R^{\sigma}{}_{\nu) \alpha \beta })  ,\nb\\
     A_{\mu\nu}^{(3)} &=& a_{3,X}  \varepsilon^{\delta\tau\alpha \beta} R_{\alpha \beta \rho \sigma} R^{\sigma}{}_{\tau} \phi^\rho \phi_\delta   \phi_{(\nu}\phi_{\mu)} \nb\\
&&    + a_3 \varepsilon^{\sigma\omega\alpha\beta} \phi_\rho \phi_\sigma R^{\rho }{}_{(\mu \alpha \beta } R_{\nu)\omega}
       -  \varepsilon^{\rho\beta\alpha}{}_{(\nu} \nabla_\sigma\nabla_\alpha  \left(a_3 \phi_{\mu)} \phi_\rho R^{\sigma}{}_{\beta} \right)  \nb\\
&&  -  \varepsilon^{\sigma\rho\alpha\beta}\nabla_\beta\nabla_\alpha \left(a_3 \phi_{(\nu} \phi_\sigma R_{\mu)\rho} \right)
       + \varepsilon^{\sigma\beta\alpha}{}_{(\nu}\nabla_\rho \nabla_\alpha \left(a_3  \phi^\rho \phi_\sigma R_{\mu)}{}_{\beta} \right)   \nb\\
&&  - \frac{1}{2}  \varepsilon^{\lambda}{}_{(\nu}{}^{\alpha\beta} \nabla_\sigma \nabla_{\mu)} \left( a_3 \phi_\rho \phi_\lambda R^{\rho\sigma }{}_{ \alpha \beta }  \right)  - \frac{1}{2}  \varepsilon^{\sigma\lambda\alpha\beta} \nabla_\lambda \nabla_{(\nu} \left( a_3 \phi_\rho \phi_\sigma  R^{\rho }{}_{\mu) \alpha \beta } \right)   \nb\\
&&  + \frac{1}{2}  \varepsilon^{\sigma}{}_{(\nu}{}^{\alpha\beta} \nabla_\lambda\nabla^\lambda \left( a_3 \phi_\rho \phi_\sigma  R^{\rho }{}_{\mu) \alpha \beta }\right) + \frac{1}{2}  \varepsilon^{\tau \lambda\alpha\beta}\nabla_\sigma \nabla_\lambda \left( a_3  \phi_\rho \phi_\tau R^{\rho\sigma }{}_{ \alpha \beta } g_{(\nu\mu)} \right) ,\nb \\
A_{\mu\nu}^{(4)} &=& a_{4,X} \varepsilon^{\delta\tau\rho\sigma} R_{\rho\sigma \alpha\beta} R^{\alpha \beta}{}_{ \delta\tau} \phi^\lambda \phi_\lambda \phi_{(\nu}\phi_{\mu)}
    + a_4  \phi_{(\mu} \phi_{\nu)} \varepsilon^{\lambda\kappa\rho\sigma} R_{\rho \sigma \alpha \beta} R^{\alpha \beta }{}_{\lambda\kappa} \nb\\
&&  +4 \varepsilon^{\alpha\rho\sigma}{}_{(\nu} \nabla_\alpha\nabla_\beta(a_4 \phi_\lambda \phi^\lambda ) R^{\beta}{}_{\mu)\rho\sigma}
      -8 \varepsilon^{\alpha\beta\rho}{}_{(\nu} \nabla_\alpha(a_4 \phi_\lambda \phi^\lambda ) \nabla_\rho R_{\beta\mu)} .
\eqn

The expression of $B_{\mu\nu}$ in Eq.(\ref{2.8}) contains seven terms,
\bqn
B_{\mu\nu}=\sum_{A=1}^7 B_{\mu\nu}^{(A)},
\eqn
where
\bqn
B_{\mu\nu}^{(1)} &=&  \varepsilon^{\tau\lambda\alpha\beta} b_{1,X} \phi_{(\nu} \phi_{\mu)} R_{\alpha\beta\rho\sigma} \phi^\rho \phi_{\tau} \phi^\sigma_\lambda
    + \varepsilon^{\sigma\kappa\alpha\beta} b_1 R^{\rho}{}_{(\nu\alpha\beta} \phi_\rho \phi_\sigma \phi_{\mu)\kappa}  \nb\\
&&  -  \varepsilon^{\rho\beta\alpha}{}_{(\nu} \nabla_\sigma\nabla_\alpha \left( b_1 \phi_{\mu)} \phi_\rho \phi^\sigma_{\beta} \right)
     + \varepsilon^{\sigma\beta\alpha}{}_{(\nu} \nabla_\rho\nabla_\alpha \left(  b_1 \phi^\rho \phi_\sigma \phi_{\mu)\beta}  \right)
     -  \varepsilon^{\sigma\rho\alpha\beta} \nabla_\beta\nabla_\alpha \left( b_1 \phi_{(\nu} \phi_\sigma \phi_{\mu)\rho} \right) \nb\\
&&  - \frac{1}{2} \left [ \varepsilon^{\lambda}{}_{(\nu}{}^{\alpha\beta} \nabla_\sigma ( b_1 R^{\rho\sigma}{}_{\alpha\beta} \phi_\lambda \phi_{\mu)} ) \phi_\rho
          + \varepsilon^{\sigma\lambda\alpha\beta} \nabla_\lambda( b_1 R^{\rho}{}_{(\mu\alpha\beta} \phi_\rho \phi_{\nu)} ) \phi_\sigma
          - \varepsilon^{\sigma}{}_{(\nu}{}^{\alpha\beta} \nabla_\lambda (b_1 R^{\rho}{}_{\mu)\alpha\beta} \phi_\rho \phi_\sigma \phi^\lambda )  \right]      ,\nb\\
B_{\mu\nu}^{(2)} &=&  \varepsilon^{\kappa\lambda\alpha\beta} b_{2,X}\phi_{(\nu} \phi_{\mu)} R_{\alpha\beta\rho\sigma} \phi^\rho_\kappa \phi^\sigma_\lambda
    + \varepsilon^{\sigma\kappa\alpha\beta} b_2 R^{\rho}{}_{(\nu\alpha\beta}  \phi_{\rho\sigma} \phi_{\mu)\kappa}  \nb\\
&&   -2 \varepsilon^{\rho\beta\alpha}{}_{(\nu} \nabla_\sigma \nabla_\alpha \left( b_2 \phi_{\mu)\rho} \phi^\sigma_{\beta} \right)
       -  \varepsilon^{\sigma\rho\alpha\beta} \nabla_\beta\nabla_\alpha \left( b_2 \phi_{(\nu\sigma} \phi_{\mu)\rho} \right)   \nb\\
&& +  \varepsilon^{\lambda\alpha\beta}{}_{(\mu} \nabla_\rho  \left( b_2  R^{\rho\sigma}{}_{\alpha\beta} \phi_{\sigma\lambda}  \phi_{\nu)} \right)
      +  \varepsilon^{\rho\lambda\alpha\beta} \nabla_\rho \left( b_2  R^{\sigma}{}_{(\mu\alpha\beta} \phi_{\sigma\lambda}  \phi_{\nu)} \right)
      +  \varepsilon^{\rho\alpha\beta}{}_{(\mu} \nabla_\lambda \left(  b_2  R^{\sigma}{}_{\nu)\alpha\beta} \phi_{\sigma\rho}  \phi^\lambda \right)      ,\nb\\
 B_{\mu\nu}^{(3)}&=&  \varepsilon^{\kappa\gamma\alpha\beta} b_{3,X} \phi_{(\mu} \phi_{\nu)} R_{\alpha\beta\rho\sigma} \phi^\sigma \phi^\rho_{\kappa} \phi^\lambda_{\gamma} \phi_\lambda
    + \varepsilon^{\sigma\kappa\alpha\beta} b_3 R^{\rho}{}_{(\nu\alpha\beta} \phi_{\mu)} \phi_{\rho\sigma} \phi^\lambda_{\kappa} \phi_\lambda
    + \varepsilon^{\gamma\lambda\alpha\beta} b_3 R^{\rho\sigma}{}_{\alpha\beta} \phi_\sigma \phi_{\rho \gamma} \phi_{(\mu \lambda} \phi_{\nu)} \nb\\
&&  - \varepsilon^{\rho\beta\alpha}{}_{(\nu} \nabla_\sigma \nabla_\alpha \left( b_3  \phi^\sigma \phi_{\mu)\rho} \phi^\lambda_{ \beta} \phi_\lambda \right)
       - \varepsilon^{\sigma\rho\alpha\beta}\nabla_\beta \nabla_\alpha \left( b_3  \phi_{(\mu} \phi_{\nu)\sigma} \phi^\lambda_{\rho} \phi_\lambda \right)
       + \varepsilon^{\sigma\beta\alpha}{}_{(\nu}\nabla_\rho \nabla_\alpha \left( b_3  \phi_{\mu)} \phi^{\rho}_{\sigma} \phi^\lambda_{\beta} \phi_\lambda \right)  \nb\\
&&  - \frac{1}{2}   \left[ \varepsilon_{(\mu}{}^{\kappa\alpha\beta} \nabla_\rho \left( b_3 R^{\rho\sigma}{}_{ \alpha\beta} \phi_\sigma  \phi_{\lambda \kappa} \phi^\lambda  \phi_{\nu)}  \right)
    + \varepsilon^{\rho\kappa\alpha\beta} \nabla_\rho \left( b_3 R_{(\mu}{}^{\sigma}{}_{ \alpha\beta} \phi_\sigma  \phi_{\lambda \kappa} \phi^\lambda  \phi_{\nu)}  \right)      \right]       \nb\\
&&  - \frac{1}{2}   \left[ \varepsilon^{\kappa}{}_{(\nu}{}^{\alpha\beta} \nabla_\lambda \left( b_3 R^{\rho\sigma}{}_{ \alpha\beta} \phi_\sigma \phi_{\rho\kappa} \phi^\lambda \phi_{\mu)}  \right)
    + \varepsilon^{\lambda\kappa\alpha\beta} \nabla_\kappa \left( b_3 R^{\rho\sigma}{}_{ \alpha\beta} \phi_\sigma \phi_{\rho\lambda} \phi_{(\mu} \phi_{\nu)} \right)    \right]       \nb\\
&&+ \frac{1}{2}   \left[ \varepsilon_{(\mu}{}^{\rho\alpha\beta} \nabla_\kappa \left( b_3 R_{\nu)}{}^{\sigma}{}_{ \alpha\beta} \phi_\sigma  \phi_{\lambda \rho} \phi^\lambda  \phi^\kappa  \right)
     +\varepsilon^{\lambda}{}_{(\nu}{}^{\alpha\beta} \nabla_\kappa \left( b_3 R^{\rho\sigma}{}_{ \alpha\beta} \phi_\sigma \phi_{\rho\lambda} \phi_{\mu)} \phi^\kappa  \right)   \right]       ,\nb\\
 B_{\mu\nu}^{(4)}&=& \varepsilon^{\kappa\gamma\alpha\beta} b_{4,X} \phi_{(\mu}\phi_{\nu)} R_{\alpha\beta\rho\sigma} \phi_{\gamma} \phi^\rho_{\kappa} \phi^\sigma_\lambda \phi^\lambda
    + \varepsilon^{\sigma\kappa\alpha\beta} b_4 R^{\rho}{}_{(\nu\alpha\beta} \phi_\kappa \phi_{\rho\sigma} \phi_{\mu)\lambda} \phi^\lambda
    + \varepsilon^{\gamma\lambda\alpha\beta} b_4 R^{\rho\sigma}{}_{\alpha\beta} \phi_\lambda \phi_{\rho \gamma} \phi_{\sigma(\nu} \phi_{\mu)}      \nb\\
&&  - \Big[ \varepsilon^{\rho\beta\alpha}{}_{(\nu} \nabla_\sigma \nabla_\alpha \left( b_4 \phi_\beta \phi_{\mu)\rho} \phi^{\sigma}_{\lambda} \phi^\lambda \right)
                    + \varepsilon^{\sigma\rho\alpha\beta} \nabla_\beta\nabla_\alpha \left( b_4\phi_\rho \phi_{(\nu\sigma} \phi_{\mu)\lambda} \phi^\lambda \right)
                    - \varepsilon^{\sigma\beta\alpha}{}_{(\nu} \nabla_\rho\nabla_\alpha \left( b_4 \phi_\beta \phi^{\rho}_{\sigma} \phi_{\mu)\lambda} \phi^\lambda \right)\Big]   \nb \\
&&  - \frac{1}{2}  \left[ \varepsilon_{(\mu}{}^{\kappa\alpha\beta} \nabla_\rho \left( b_4 R^{\rho\sigma}{}_{\alpha\beta} \phi_\kappa \phi_{\sigma\lambda} \phi^\lambda \phi_{\nu)} \right)
                    + \varepsilon^{\rho\kappa\alpha\beta} \nabla_\rho \left( b_4 R_{(\mu}{}^{\sigma}{}_{\alpha\beta} \phi_\kappa \phi_{\sigma\lambda} \phi^\lambda \phi_{\nu)} \right)     \right]    \nb\\
&&  - \frac{1}{2}  \left[ \varepsilon^{\kappa\lambda\alpha\beta} \nabla_\sigma \left( b_4 R^{\rho\sigma}{}_{\alpha\beta} \phi_\lambda \phi_{\rho\kappa} \phi_{(\nu} \phi_{\mu)} \right)
                    + \varepsilon^{\sigma\kappa\alpha\beta} \nabla_\lambda \left( b_4 R^{\rho}{}_{(\mu\alpha\beta} \phi_\kappa \phi_{\rho\sigma} \phi^\lambda \phi_{\nu)} \right)    \right]    \nb\\
&&     + \frac{1}{2}  \left[ \varepsilon_{(\mu}{}^{\rho\alpha\beta} \nabla_\kappa \left( b_4 R_{\nu)}{}^{\sigma}{}_{\alpha\beta} \phi_\rho \phi_{\sigma\lambda} \phi^\lambda \phi^\kappa \right)
                    + \varepsilon^{\sigma\lambda\alpha\beta} \nabla_\kappa \left( b_4 R^{\rho}{}_{(\mu\alpha\beta} \phi_\lambda \phi_{\rho\sigma} \phi_{\nu)} \phi^\kappa \right)    \right]     ,\nb\\
 B_{\mu\nu}^{(5)}&=& \varepsilon^{\kappa\gamma\alpha\beta} b_{5,X} \phi_{(\mu}\phi_{\nu)} R_{\alpha\rho\sigma\lambda} \phi^\rho \phi_\beta \phi^\sigma_{\kappa} \phi^\lambda_{\gamma}
    - 2\varepsilon^{\tau\sigma\alpha\beta} b_5 R^{\rho}{}_{\alpha(\nu\lambda} \phi_\rho \phi_\beta \phi_{\mu) \tau } \phi^\lambda_{\sigma}   \nb\\
&&  +\varepsilon^{\rho\lambda\alpha\beta}\nabla_\alpha\nabla_\sigma \left( b_5 \phi_{(\mu} \phi_\beta \phi^\sigma_{ \rho} \phi_{\nu)\lambda} \right)
                    + \varepsilon^{\alpha\rho}{}_{(\mu}{}^{\beta}\nabla_\lambda\nabla_\sigma \left( b_5 \phi_{\nu)} \phi_\beta \phi^\sigma_{ \alpha} \phi^\lambda_{ \rho} \right) \nb\\
&&                - \varepsilon^{\alpha\lambda}{}_{(\mu}{}^{\beta}\nabla_\rho\nabla_\sigma \left( b_5 \phi^\rho \phi_\beta \phi^\sigma_{\alpha} \phi_{ \nu)\lambda} \right)
    +  \varepsilon_{(\mu}{}^{\kappa\alpha\beta} \nabla_\sigma \left( b_5 R^{\rho}{}_{\alpha}{}^{\sigma\lambda} \phi_\rho \phi_\beta \phi_{\lambda \kappa} \phi_{\nu)}  \right)     \nb\\
&&+ \varepsilon^{\sigma\kappa\alpha\beta} \nabla_\sigma \left( b_5 R^{\rho}{}_{\alpha(\mu}{}^{\lambda} \phi_\rho \phi_\beta \phi_{\lambda \kappa} \phi_{\nu)}  \right)
    - \varepsilon_{(\mu}{}^{\sigma\alpha\beta} \nabla_\kappa \left( b_5 R^{\rho}{}_{\alpha\nu)}{}^{\lambda} \phi_\rho \phi_\beta \phi_{\lambda \sigma} \phi^\kappa\right)          ,  \nb\\
 B_{\mu\nu}^{(6)}&=& \ \varepsilon^{\kappa\tau\alpha\beta} b_{6,X} \phi_{(\mu}\phi_{\nu)}  R_{\beta\gamma} \phi_\alpha \phi^{\gamma}_{\kappa} \phi^{\lambda} _{\tau} \phi_\lambda
 +  \varepsilon^{\gamma \sigma\alpha\beta} b_6 R_{\beta(\mu} \phi_\alpha \phi_{\gamma\nu)} \phi^{\lambda}_{\sigma} \phi_\lambda
    + \varepsilon^{\lambda\rho\alpha\beta} b_6 R_{\beta\gamma} \phi_\alpha \phi^{\gamma}_{\lambda} \phi_{\rho(\nu} \phi_{\mu)}  \nb\\
 &&   - \frac{1}{2} \varepsilon^{\tau\gamma\alpha\beta} \nabla_\beta \nabla_{(\mu} \left( b_6 g^{\lambda\rho} \phi_\alpha \phi_{\nu)\tau} \phi_{\rho\gamma} \phi_\lambda  \right)
  - \frac{1}{2} \varepsilon^{\beta\tau\alpha}{}_{(\mu}  \nabla_\gamma \nabla_{\nu)} \left( b_6 g^{\gamma \sigma} g^{\lambda\rho} \phi_\alpha \phi_{\sigma\beta} \phi_{\rho\tau} \phi_\lambda \right)   \nb\\
 &&     + \frac{1}{2} \varepsilon^{\beta\gamma\alpha}{}_{(\mu} \nabla_\tau \nabla_\kappa \left( b_6  g^{\lambda\rho} \phi_\alpha \phi_{\nu)\beta} \phi_{\rho\gamma} \phi_\lambda g^{\kappa\tau}  \right)
 + \frac{1}{2} \varepsilon^{\tau\kappa\alpha\beta} \nabla_\beta \nabla_\gamma \left(  b_6 g^{\gamma \sigma} g^{\lambda\rho} \phi_\alpha \phi_{\sigma\tau} \phi_{\rho\kappa} \phi_\lambda  g_{(\nu\mu)} \right) \nb\\
&&   + \frac{1}{2} \varepsilon^{\beta\tau\alpha}{}_{(\mu}  \nabla_{\nu)} \nabla_\gamma \left( b_6 g^{\gamma \sigma} g^{\lambda\rho} \phi_\alpha \phi_{\sigma\beta} \phi_{\rho\tau} \phi_\lambda  \right)
  - \frac{1}{2}  \varepsilon^{\beta\kappa\alpha}{}_{(\mu} \nabla_{\nu)} \nabla_\gamma \left( b_6 g^{\gamma \sigma} g^{\lambda\rho} \phi_\alpha \phi_{\sigma\beta} \phi_{\rho\kappa} \phi_\lambda \right)  \nb\\
&&  -\frac{1}{2} \varepsilon^{\xi\sigma\alpha\beta}  \nabla_\xi \left( b_6 R_{\beta(\nu} \phi_\alpha g^{\lambda\rho} \phi_{\sigma\rho} \phi_\lambda \phi_{\mu)} \right)
      -\frac{1}{2} \varepsilon_{(\mu}{}^{\xi\alpha\beta}  \nabla_\sigma \left( b_6 R_{\beta\gamma} \phi_\alpha g^{\gamma \sigma} g^{\lambda\rho} \phi_{\xi\rho} \phi_\lambda \phi_{\nu)} \right)   \nb\\
&&  +\frac{1}{2} \varepsilon_{(\mu}{}^{\sigma\alpha\beta}  \nabla_\xi \left( b_6 R_{\beta\nu)} \phi_\alpha g^{\lambda\rho} \phi_{\sigma\rho} \phi_\lambda \phi_\tau g^{\tau\xi} \right)
       -\frac{1}{2} \varepsilon^{\xi\rho\alpha\beta}\nabla_\rho \left(  b_6 R_{\beta\gamma} \phi_\alpha g^{\gamma \sigma} \phi_{\xi\sigma} \phi_{(\nu}\phi_{\mu)}  \right)  \nb\\
&&   -\frac{1}{2} \varepsilon^{\xi}{}_{(\nu}{}^{\alpha\beta}  \nabla_\rho  \left( b_6 R_{\beta\gamma} \phi_\alpha g^{\gamma \sigma} g^{\lambda\rho}\phi_{\xi\sigma} \phi_\lambda  \phi_{\mu)}  \right)
        +\frac{1}{2} \varepsilon^{\rho}{}_{(\nu}{}^{\alpha\beta}  \nabla_\xi \left( b_6 R_{\beta\gamma} \phi_\alpha g^{\gamma \sigma} \phi_{\rho\sigma} \phi_{\mu)}  \phi_\tau g^{\tau\xi}  \right)   \nb\\
 B_{\mu\nu}^{(7)}&=&   \varepsilon^{\tau\gamma\alpha\beta} b_{7,X} \phi_{(\mu}\phi_{\nu)}  R^{\rho}{}_{\sigma\alpha\beta} \phi_\rho \phi_{\tau} \phi_{\kappa \gamma} g^{\sigma\kappa} g^{\lambda\delta} \phi_{\lambda\delta}
 - \varepsilon^{\xi\beta\alpha}{}_{(\nu}  \nabla_\sigma \nabla_\alpha \left(  b_7 \phi_{\mu)} \phi_\xi \phi_{\kappa\beta} g^{\sigma\kappa} g^{\lambda\delta} \phi_{\lambda\delta} \right)  \nb\\
&& -  \varepsilon^{\sigma\xi\alpha\beta}  \nabla_\beta \nabla_\alpha \left( b_7 \phi_{(\nu} \phi_\sigma \phi_{\mu)\xi}  g^{\lambda\delta} \phi_{\lambda\delta} \right)
 +  \varepsilon^{\sigma\beta\alpha}{}_{(\nu}  \nabla_\xi \nabla_\alpha \left( b_7 \phi_\rho \phi_\sigma \phi_{\mu)\beta} g^{\lambda\delta} \phi_{\lambda\delta} g^{\rho \xi} \right) \nb\\
 && + \varepsilon^{\sigma\kappa\alpha\beta} b_7 R^{\rho}{}_{(\mu\alpha\beta} \phi_\rho \phi_\sigma \phi_{\nu)\kappa} g^{\lambda\delta} \phi_{\lambda\delta}
  + \varepsilon^{\lambda\tau\alpha\beta} b_7 R^{\rho}{}_{\sigma\alpha\beta} \phi_\rho \phi_\lambda \phi_{\kappa\tau} g^{\sigma\kappa} \phi_{(\mu\nu)}   \nb\\
&&  -\frac{1}{2}  \varepsilon^{\xi}{}_{(\nu}{}^{\alpha\beta} \nabla_\kappa \left( b_7 R^{\rho}{}_{\sigma\alpha\beta} \phi_\rho \phi_\xi  g^{\sigma\kappa} g^{\lambda\delta} \phi_{\lambda\delta} \phi_{\mu)} \right)
  -\frac{1}{2}  \varepsilon^{\kappa\xi\alpha\beta} \nabla_\xi \left(  b_7 R^{\rho}{}_{(\mu\alpha\beta} \phi_\rho \phi_\kappa g^{\lambda\delta} \phi_{\lambda\delta} \phi_{\nu)} \right)  \nb\\
&&  +\frac{1}{2}  \varepsilon^{\kappa}{}_{(\nu}{}^{\alpha\beta} \nabla_\xi \left( b_7 R^{\rho}{}_{\mu)\alpha\beta} \phi_\rho \phi_\kappa  g^{\lambda\delta} \phi_{\lambda\delta} \phi_\gamma g^{\gamma \xi}  \right)
   - \frac{1}{2} \varepsilon^{\xi\delta\alpha\beta} \nabla_{(\nu} \left( b_7 R^{\rho}{}_{\sigma\alpha\beta} \phi_\rho \phi_\xi \phi_{\kappa\delta} g^{\sigma\kappa} \phi_{\mu)} \right)  \nb\\
&&   - \frac{1}{2} \varepsilon^{\lambda\xi\alpha\beta} \nabla_{(\mu} \left(  b_7 R^{\rho}{}_{\sigma\alpha\beta} \phi_\rho \phi_\lambda \phi_{\kappa\xi} g^{\sigma\kappa} \phi_{\nu)} \right)
  + \frac{1}{2} \varepsilon^{\lambda\delta\alpha\beta} \nabla_\xi \left( b_7 R^{\rho}{}_{\sigma\alpha\beta} \phi_\rho \phi_\lambda \phi_{\kappa\delta} g^{\sigma\kappa} g_{(\mu\nu)} \phi_\gamma g^{\gamma \xi} \right)  .
 \eqn

The energy-momentum tensor of scalar field $\phi$ in Eq.(\ref{2.8}) is given by
\bqn\lb{sft}
T^{\phi}_{\mu\nu}=  \frac{1}{2}g_{\mu\nu} g^{\rho\sigma} \partial_\rho \phi\partial_\sigma \phi + g_{\mu\nu}V(\phi)
    -\partial_\mu \phi\partial_\nu \phi  .
\eqn

The expression of $F_{,\phi}$ in Eq.(\ref{2.12}) is given by
\bqn
F_{,\phi} = \sum_{A=1}^4 F_{a_A} + \sum_{B=1}^4 F_{b_B},
\eqn
where $F_{a_A}$ and $F_{b_B}$ are
\bqn
F_{a_1}
&=&   a_{1,\phi}\varepsilon^{\mu\nu\alpha \beta} R_{\alpha \beta \rho}{}^{\sigma} R_{\mu \nu}{}^{\rho\lambda}\phi_\sigma \phi_\lambda
      - 2\varepsilon^{\mu\nu\alpha \beta}\nabla_\gamma\left( a_{1,X} R_{\alpha \beta \rho}{}^{\sigma} R_{\mu \nu}{}^{\rho\lambda}\phi_\sigma \phi_\lambda\phi^\gamma \right)  \nb\\
&&    -  \varepsilon^{\mu\nu\alpha \beta}\nabla_\sigma\left(a_1 R_{\alpha \beta \rho}{}^{\sigma} R_{\mu \nu}{}^{\rho\lambda}\phi_\lambda \right)
      -  \varepsilon^{\mu\nu\alpha \beta}\nabla_\lambda\left(a_1 R_{\alpha \beta \rho}{}^{\sigma} R_{\mu \nu}{}^{\rho\lambda}\phi_\sigma \right)  , \nb\\
F_{a_2}
&=&    a_{2,\phi}  \varepsilon^{\mu\nu\alpha \beta} R_{\alpha \beta \rho \sigma} R_{\mu}{}^{ \lambda \rho \sigma}\phi_\sigma \phi_\lambda
      - 2 \varepsilon^{\mu\nu\alpha \beta} \nabla_\gamma\left( a_{2,X} R_{\alpha \beta \rho \sigma} R_{\mu}{}^{ \lambda \rho \sigma}\phi_\nu \phi_\lambda\phi^\gamma \right)  \nb\\
&&    -   \varepsilon^{\mu\nu\alpha \beta} \nabla_\nu\left(a_2 R_{\alpha \beta \rho \sigma} R_{\mu}{}^{ \lambda \rho \sigma}\phi_\lambda \right)
      -   \varepsilon^{\mu\nu\alpha \beta}\nabla_\lambda\left(a_2 R_{\alpha \beta \rho \sigma} R_{\mu}{}^{ \lambda \rho \sigma}\phi_\nu \right)   , \nb\\
F_{a_3}
&=&    a_{3,\phi} \varepsilon^{\mu\nu\alpha \beta} R_{\alpha \beta}{}^{ \rho}{}_{ \sigma} R^{\sigma}{}_{\nu}\phi_\rho \phi_\mu
      - 2  \varepsilon^{\mu\nu\alpha \beta} \nabla_\gamma \left( a_{3,X} R_{\alpha \beta}{}^{ \rho}{}_{ \sigma} R^{\sigma}{}_{\nu}\phi_\rho \phi_\mu\phi^\gamma \right)  \nb\\
&&    -  \varepsilon^{\mu\nu\alpha \beta} \nabla_\rho\left(a_3 R_{\alpha \beta}{}^{ \rho}{}_{ \sigma} R^{\sigma}{}_{\nu}\phi_\mu \right)
      -  \varepsilon^{\mu\nu\alpha \beta} \nabla_\mu\left(a_3 R_{\alpha \beta}{}^{ \rho}{}_{ \sigma} R^{\sigma}{}_{\nu}\phi_\rho \right)    , \nb\\
F_{a_4}
&=&   a_{4,\phi} \varepsilon^{\mu\nu\alpha \beta} R_{\rho\sigma \alpha \beta} R^{\alpha\beta}{}_{\mu \nu}\phi^\lambda \phi_\lambda
      - 2   \varepsilon^{\mu\nu\alpha \beta} \nabla_\gamma \left( a_{4,X} R_{\rho\sigma \alpha \beta} R^{\alpha\beta}{}_{\mu \nu}\phi^\lambda \phi_\lambda\phi^\gamma \right)  \nb\\
&&    - 2  \varepsilon^{\mu\nu\alpha \beta} \nabla_\lambda\left(a_4 R_{\rho\sigma \alpha \beta} R^{\alpha\beta}{}_{\mu \nu}\phi^\lambda \right)    , \nb\\
F_{b_1}
&=&    b_{1,\phi} \varepsilon^{\mu\nu\alpha \beta} R_{\alpha \beta}{}^{\rho\sigma } \phi_\rho\phi_\mu\phi_{\sigma\nu }
                -\varepsilon^{\mu\nu\alpha \beta} \nabla_\lambda\left( 2b_{1,X} R_{\alpha \beta}{}^{\rho\sigma } \phi^\lambda \phi_\rho\phi_\mu\phi_{\sigma\nu }\right)  \nb\\
&&           -\varepsilon^{\mu\nu\alpha \beta} \nabla_\rho \left(b_1 R_{\alpha \beta}{}^{\rho\sigma } \phi_\mu\phi_{\sigma\nu }\right)
               -\varepsilon^{\mu\nu\alpha \beta} \nabla_\mu \left(b_1 R_{\alpha \beta}{}^{\rho\sigma } \phi_\rho\phi_{\sigma\nu }\right) \nb\\
&&                +\varepsilon^{\mu\nu\alpha \beta} \nabla_\nu\nabla_\sigma \left(b_1 R_{\alpha \beta}{}^{\rho\sigma }\phi_\rho\phi_\mu \right)
                   , \nb\\
F_{b_2}
&=&    b_{2,\phi} \varepsilon^{\mu\nu\alpha \beta} R_{\alpha \beta}{}^{\rho\sigma } \phi_{\rho\mu}\phi_{\sigma\nu }
                                     - \varepsilon^{\mu\nu\alpha \beta} \nabla_\gamma \left( 2b_{2,X} R_{\alpha \beta}{}^{\rho\sigma } \phi^\gamma \phi_{\rho\mu}\phi_{\sigma\nu }\right) \nb\\
&&                                   + \varepsilon^{\mu\nu\alpha\beta} \nabla_\mu\nabla_\rho\left(b_2  R_{\alpha \beta}{}^{\rho\sigma }  \phi_{\sigma\nu } \right)
                                     + \varepsilon^{\mu\nu\alpha \beta} \nabla_\nu \nabla_\sigma\left(b_2 R_{\alpha \beta}{}^{\rho\sigma }  \phi_{\rho\mu } \right)
                                       , \nb\\
F_{b_3}
&=&    b_{3,\phi} \varepsilon^{\mu\nu \alpha \beta} R_{\alpha \beta \rho\sigma}\phi_\sigma \phi_{\rho\mu} \phi_{\lambda\nu} \phi^\lambda
                - 2 \varepsilon^{\mu\nu \alpha \beta} \nabla_\gamma \left( b_{3,\phi} R_{\alpha \beta \rho\sigma}\phi^\gamma \phi_\sigma \phi_{\rho\mu} \phi_{\lambda\nu} \phi^\lambda \right) \nb\\
&&              -  \varepsilon^{\mu\nu \alpha \beta} \nabla_\sigma \left( b_3 R_{\alpha \beta \rho\sigma} \phi_{\rho\mu} \phi_{\lambda\nu} \phi^\lambda \right)
                -  \varepsilon^{\mu\nu \alpha \beta} \nabla^\lambda\left( b_3 R_{\alpha \beta \rho\sigma} \phi_\sigma \phi_{\rho\mu} \phi_{\lambda\nu} \right) \nb\\
&&              +  \varepsilon^{\mu\nu \alpha \beta} \nabla_\mu\nabla_\rho\left( b_3 R_{\alpha \beta \rho\sigma} \phi_\sigma\phi^\lambda\phi_{\lambda\nu} \right)
                + \varepsilon^{\mu\nu \alpha \beta} \nabla_\nu\nabla_\lambda\left( b_3 R_{\alpha \beta \rho\sigma} \phi_\sigma \phi^\lambda \phi_{\rho\mu} \right)
                 , \nb\\
F_{b_4}
&=&     b_{4,\phi} \varepsilon^{\mu\nu \alpha \beta} R_{\alpha \beta \rho\sigma} \phi_\nu \phi_{\rho\mu } \phi_{\sigma\lambda} \phi^\lambda
                - 2 \varepsilon^{\mu\nu \alpha \beta} \nabla_\gamma \left( b_{4,\phi} R_{\alpha \beta \rho\sigma}\phi^\gamma \phi_\nu \phi_{\rho\mu } \phi_{\sigma\lambda} \phi^\lambda \right) \nb\\
&&              - \varepsilon^{\mu\nu \alpha \beta} \nabla_\nu \left( b_4 R_{\alpha \beta \rho\sigma} \phi_{\rho\mu} \phi_{\sigma\lambda} \phi^\lambda \right)
                -  \varepsilon^{\mu\nu \alpha \beta} \nabla_\lambda\left( b_4 R_{\alpha \beta \rho\sigma} \phi_\nu \phi_{\rho\mu} \phi_{\sigma}^{\lambda}  \right) \nb\\
&&              +  \varepsilon^{\mu\nu \alpha \beta} \nabla_\mu\nabla_\rho\left( b_4 R_{\alpha \beta \rho\sigma} \phi_\nu\phi^\lambda\phi_{\sigma\lambda} \right)
                +  \varepsilon^{\mu\nu \alpha \beta} \nabla_\lambda\nabla_\sigma\left( b_4 R_{\alpha \beta \rho\sigma} \phi_\nu \phi^\lambda \phi_{\rho\mu} \right)
                 , \nb\\
F_{b_5}
&=&   b_{5,\phi} \varepsilon^{\mu\nu \alpha \beta} R_{\alpha}{}^{ \rho\sigma \lambda }\phi_\rho \phi_\beta \phi_{\sigma\mu} \phi_{\lambda\nu}
                - 2 \varepsilon^{\mu\nu \alpha \beta} \nabla_\gamma \left( b_{5,\phi} R_{\alpha}{}^{ \rho\sigma \lambda }\phi^\gamma \phi_\rho \phi_\beta \phi_{\sigma\mu} \phi_{\lambda\nu} \right) \nb\\
&&              -  \varepsilon^{\mu\nu \alpha \beta} \nabla_\rho \left( b_5 R_{\alpha}{}^{ \rho\sigma \lambda } \phi_\beta \phi_{\sigma\mu} \phi_{\lambda\nu} \right)
                -  \varepsilon^{\mu\nu \alpha \beta} \nabla_\beta\left( b_5 R_{\alpha}{}^{ \rho\sigma \lambda } \phi_\rho \phi_{\sigma\mu} \phi_{\lambda\nu} \right) \nb\\
&&              +  \varepsilon^{\mu\nu \alpha\beta} \nabla_\mu\nabla_\sigma\left( b_5  R_{\alpha}{}^{ \rho\sigma \lambda } \phi_\rho \phi_\beta \phi_{\lambda\nu} \right)
                +  \varepsilon^{\mu\nu \alpha\beta} \nabla_\nu\nabla_\lambda\left( b_5 R_{\alpha}{}^{ \rho\sigma \lambda } \phi_\rho \phi_\beta \phi_{\sigma\mu} \right)
                , \nb\\
F_{b_6}
&=&    b_{6,\phi} \varepsilon^{\mu\nu \alpha \beta} R_{\beta}{}^{ \gamma} \phi_\alpha \phi_{\gamma\mu} \phi_{\lambda\nu} \phi^\lambda
                - 2 \varepsilon^{\mu\nu \alpha \beta} \nabla_\tau \left( b_{6,\phi} R_{\beta}{}^{ \gamma} \phi^\tau \phi_\alpha \phi_{\gamma\mu} \phi_{\lambda\nu} \phi^\lambda \right) \nb\\
&&              -  \varepsilon^{\mu\nu \alpha \beta} \nabla_\alpha \left( b_6 R_{\beta}{}^{ \gamma} \phi_{\gamma\mu} \phi_{\lambda\nu} \phi^\lambda \right)
                -  \varepsilon^{\mu\nu \alpha \beta} \nabla_\lambda\left( b_6 R_{\beta}{}^{ \gamma} \phi_\alpha \phi_{\gamma\mu} \phi^{\lambda}_{\nu} \right) \nb\\
&&              + \varepsilon^{\mu\nu \alpha \beta} \nabla_\mu\nabla_\gamma\left( b_6 R_{\beta}{}^{ \gamma} \phi_\alpha \phi_{\lambda\nu} \phi^\lambda \right)
                +  \varepsilon^{\mu\nu \alpha \beta} \nabla_\nu\nabla_\lambda\left( b_6 R_{\beta}{}^{ \gamma} \phi_\alpha \phi_{\gamma\mu} \phi^\lambda \right)
                  , \nb\\
F_{b_7}
&=&    b_{7,\phi}  \varepsilon^{\mu\nu\alpha \beta} R_{\alpha \beta}{}^{\rho\sigma } \phi_\rho\phi_\mu\phi_{\sigma\nu } \nabla^2\phi
                - 2 \varepsilon^{\mu\nu\alpha \beta} \nabla_\lambda\left( b_{7,X} R_{\alpha \beta}{}^{\rho\sigma } \phi^\lambda \phi_\rho\phi_\mu\phi_{\sigma\nu }\nabla^2\phi\right)  \nb\\
&&              -  \varepsilon^{\mu\nu\alpha \beta} \nabla_\rho \left( b_7 R_{\alpha \beta}{}^{\rho\sigma } \phi_\mu\phi_{\sigma\nu }\nabla^2\phi \right)
                -  \varepsilon^{\mu\nu\alpha \beta} \nabla_\mu \left( b_7 R_{\alpha \beta}{}^{\rho\sigma } \phi_\rho\phi_{\sigma\nu }\nabla^2\phi \right)  \nb\\
&&              +  \varepsilon^{\mu\nu\alpha \beta} \nabla_\nu\nabla_\sigma \left( b_7 R_{\alpha \beta}{}^{\rho\sigma }\phi_\rho\phi_\mu \nabla^2\phi \right)
                + \varepsilon^{\mu\nu\alpha \beta} \nabla^2 \left( b_7 R_{\alpha \beta}{}^{\rho\sigma } \phi_\rho\phi_\mu \phi_{\sigma\nu } \right)   .
\eqn

\section*{Appendix B: The propagation equation of gravitational wave in ghost-free parity-violating gravities }
\renewcommand{\theequation}{B.\arabic{equation}} \setcounter{equation}{0}

In Sec. \ref{sec2}, we obtain the metric field equation of motion and the scalar field equation of motion from the variation of action with respect to the metric field and the scalar field. In order to cross-check of these results, in this Appendix, we expand the field equation in weak fields and derive the propagation equation of gravitational waves in the Friedmann-Robertson-Walker universe. Note that, in the previous works \cite{TZ2019,Qiao:2019hkz}, we first carried out the cosmological perturbation expansion of the action about the spatial metric, and then obtained the propagation equation of gravitational wave by the variation of the action to the spatial perturbation metric $h_{ij}$. The comparison of these two results is helpful to validate the metric field equation of motion derived in this article.

Let us consider the spatial metric in the flat Friedmann-Robertson-Walker universe, which is written as
\bqn
g_{ij} = a^2 (\delta_{ij} + h_{ij}),
\eqn
where $a$ is the scale factor of the universe. We substitute this expansion of metric into all tensors in Appendix A, and obtain
\bqn
A^{(1)}_{ij}&=&  \varepsilon^{0kl}{}_{i}  \left( 2 a'_1 \phi'^2 \frac{1}{a^2} + 4 a_1 \phi' \phi'' \frac{1}{a^2} - 8 a_1 \phi'^2 \frac{a'}{a^3} \right) h^{\prime\prime }_{jk,l} \nb\\
                 &&  + \varepsilon^{0kl}{}_{i}  \Bigg( 2 a''_1 \phi'^2 \frac{1}{a^2} + 8 a'_1 \phi' \phi'' \frac{1}{a^2} - 12 a'_1 \phi'^2 \frac{a'}{a^3} + 4 a_1 \phi' \phi''' \frac{1}{a^2}  \nb\\
                 &&  + 4 a_1 \phi''^2 \frac{1}{a^2} - 24 a_1 \phi' \phi'' \frac{a'}{a^3} - 8 a_1 \phi'^2 \frac{a''}{a^3} + 24 a_1 \phi'^2 \frac{a'^2}{a^4} \Bigg) h^{\prime}_{jk,l} ,\nb\\
A^{(2)}_{ij}&=&  \varepsilon^{0kl}{}_{i}  \left( a'_2 \phi'^2 \frac{1}{a^2} + 2 a_2 \phi' \phi'' \frac{1}{a^2} - 2 a_2 \phi'^2 \frac{a'}{a^3} \right) h^{\prime\prime }_{jk,l} \nb\\
                 &&   -\varepsilon^{0kl}{}_{i}  \left( a'_2 \phi'^2 \frac{1}{a^2} + 2 a_2 \phi' \phi'' \frac{1}{a^2} - 2 a_2 \phi'^2 \frac{a'}{a^3} \right) h_{jk,l,m}{}^{,m} \nb\\
                 &&  + \varepsilon^{0kl}{}_{i}  \Bigg( a''_2 \phi'^2 \frac{1}{a^2} + 4 a'_2 \phi' \phi'' \frac{1}{a^2} - 4 a'_2 \phi'^2 \frac{a'}{a^3} + 2 a_2 \phi' \phi''' \frac{1}{a^2}  \nb\\
                 &&  + 2 a_2 \phi''^2 \frac{1}{a^2} - 8 a_2 \phi' \phi'' \frac{a'}{a^3} - 2 a_2 \phi'^2 \frac{a''}{a^3} + 6 a_2 \phi'^2 \frac{a'^2}{a^4} \Bigg) h^{\prime}_{jk,l} ,\nb\\
A^{(3)}_{ij}&=&  \varepsilon^{0kl}{}_{i}  \left( \frac{1}{2}a'_3 \phi'^2 \frac{1}{a^2} + a_3 \phi' \phi'' \frac{1}{a^2} - 3 a_3 \phi'^2 \frac{a'}{a^3} \right) h^{\prime\prime }_{jk,l} \nb\\
                 &&   +\varepsilon^{0kl}{}_{i}  \left( \frac{1}{2}a'_3 \phi'^2 \frac{1}{a^2} + a_3 \phi' \phi'' \frac{1}{a^2} - a_3 \phi'^2 \frac{a'}{a^3} \right) h_{jk,l,m}{}^{,m} \nb\\
                 &&  + \varepsilon^{0kl}{}_{i}  \Bigg( \frac{1}{2} a''_3 \phi'^2 \frac{1}{a^2} + 2 a'_3 \phi' \phi'' \frac{1}{a^2} - 4 a'_3 \phi'^2 \frac{a'}{a^3} + a_3 \phi' \phi''' \frac{1}{a^2}  \nb\\
                 &&  + a_3 \phi''^2 \frac{1}{a^2} - 8 a_3 \phi' \phi'' \frac{a'}{a^3} - 3 a_3 \phi'^2 \frac{a''}{a^3} + 9 a_3 \phi'^2 \frac{a'^2}{a^4} \Bigg) h^{\prime}_{jk,l} ,\nb\\
A^{(4)}_{ij}&=&  \varepsilon^{0kl}{}_{i}  \left( 4 a'_4 \phi'^2 \frac{1}{a^2} + 8 a_4 \phi' \phi'' \frac{1}{a^2} - 8 a_4 \phi'^2 \frac{a'}{a^3} \right) h''_{jk,l} \nb\\
                 &&   -\varepsilon^{0kl}{}_{i}  \left( 4 a'_4 \phi'^2 \frac{1}{a^2} + 8 a_4 \phi' \phi'' \frac{1}{a^2} - 8 a_4 \phi'^2 \frac{a'}{a^3} \right) h_{jk,l,m}{}^{,m} \nb\\
                 &&  +\varepsilon^{0kl}{}_{i}  \Bigg( 4 a''_4 \phi'^2 \frac{1}{a^2} + 16 a'_4 \phi' \phi'' \frac{1}{a^2} - 16 a'_4 \phi'^2 \frac{a'}{a^3} + 8 a_4 \phi' \phi''' \frac{1}{a^2}  \nb\\
                 &&  + 8 a_4 \phi''^2 \frac{1}{a^2} - 32 a_4 \phi' \phi'' \frac{a'}{a^3} - 8 a_4 \phi'^2 \frac{a''}{a^3} + 24 a_4 \phi'^2 \frac{a'^2}{a^4} \Bigg) h'_{jk,l} ,\nb\\
B^{(1)}_{ij}&=&  \varepsilon^{0kl}{}_{i}  \left(  b_1 \phi'^3 \frac{1}{a^2}   \right)   h''_{jk,l} \nb\\
                 &&   +\varepsilon^{0kl}{}_{i}  \left(  b'_1 \phi'^3 \frac{1}{a^2} + 3 b_1 \phi'^2 \phi'' \frac{1}{a^2} - 2 b_1 \phi'^3 \frac{a'}{a^3} \right) h'_{jk,l} ,\nb\\
B^{(2)}_{ij}&=&  \varepsilon^{0kl}{}_{i}  \left(  2 b_2 \phi' \phi'' \frac{1}{a^2} - 4 b_2 \phi'^2 \frac{a'}{a^3}   \right)   h''_{jk,l} \nb\\
                 &&   +\varepsilon^{0kl}{}_{i}  \Bigg(  2 b'_2 \phi' \phi'' \frac{1}{a^2} - 4 b'_2 \phi'^2 \frac{a'}{a^3} + 2 b_2 \phi' \phi''' \frac{1}{a^2} + 2 b_2 \phi''^2 \frac{1}{a^2} \nb\\
                 &&  + 12 b_2 \phi'^2 \frac{a'^2}{a^4} -4 b_2 \phi'^2 \frac{a''}{a^3} - 12 b_2 \phi' \phi'' \frac{a'}{a^3} \Bigg) h'_{jk,l} ,\nb\\
B^{(3)}_{ij}&=&  \varepsilon^{0kl}{}_{i}  \left( - b_3 \phi'^3 \phi'' \frac{1}{a^4} + b_3 \phi'^4 \frac{a'}{a^5}   \right)   h''_{jk,l} \nb\\
                 &&   +\varepsilon^{0kl}{}_{i}  \Bigg(  -  b'_3 \phi'^3 \phi'' \frac{1}{a^4} + b'_3 \phi'^4 \frac{a'}{a^5} - 3 \phi'^2 \phi''^2 \frac{1}{a^4} - b_3 \phi'^3 \phi''' \frac{1}{a^4} \nb\\
                 &&  -5 b_3 \phi'^4 \frac{a'^2}{a^6} + 8 b_3 \phi'^3 \phi'' \frac{a'}{a^5} +  b_3 \phi'^4 \frac{a''}{a^5} \Bigg) h'_{jk,l} ,\nb\\
B^{(4)}_{ij}&=&  \varepsilon^{0kl}{}_{i}  \left( b_4 \phi'^4 \frac{a'}{a^5} - b_4 \phi'^3 \phi'' \frac{1}{a^4}    \right)   h''_{jk,l} \nb\\
                 &&   +\varepsilon^{0kl}{}_{i}  \Bigg(  - b'_4 \phi'^3 \phi'' \frac{1}{a^4} + b'_4 \phi'^4 \frac{a'}{a^5} - 3 b_4 \phi'^2 \phi''^2 \frac{1}{a^4} -  b_4 \phi'^3 \phi''' \frac{1}{a^4} \nb\\
                 &&  - 5 b_4 \phi'^4 \frac{a'^2}{a^6} +8 b_4 \phi'^3 \phi'' \frac{a'}{a^5} + b_4 \phi'^4 \frac{a''}{a^5} \Bigg) h'_{jk,l} ,\nb\\
B^{(5)}_{ij}&=&  \varepsilon^{0kl}{}_{i}  \left( b_5 \phi'^4 \frac{a'}{a^5}  \right)   h''_{jk,l} \nb\\
                 &&   +\varepsilon^{0kl}{}_{i}  \Bigg(  b'_5 \phi'^4 \frac{a'}{a^5}  +  4 b_5 \phi'^3 \phi'' \frac{a'}{a^5}  + b_5 \phi'^4 \frac{a''}{a^5} - \frac{5}{2} b_5 \phi'^4 \frac{a'^2}{a^6} \Bigg) h'_{jk,l} ,\nb\\
B^{(7)}_{ij}&=&  \varepsilon^{0kl}{}_{i}  \left( - b_7 \phi'^3\phi'' \frac{1}{a^4} - 2 b_7 \phi'^4 \frac{a'}{a^5}  \right)   h''_{jk,l} \nb\\
                 &&   +\varepsilon^{0kl}{}_{i}  \Bigg(  - b'_7 \phi'^3\phi'' \frac{1}{a^4} - 2 b'_7 \phi'^4 \frac{a'}{a^5} - 3 b_7 \phi'^2 \phi''^2 \frac{1}{a^4} - b_7 \phi'^3 \phi''' \frac{1}{a^4} \nb\\
                 && - 4b_7 \phi'^3 \phi'' \frac{a'}{a^5} + 10 b_7 \phi'^4 \frac{a'^2}{a^6} -2 b_7 \phi'^4 \frac{a''}{a^5} \Bigg) h'_{jk,l} .
\eqn
Substituting these expressions into field equation in Eq.(\ref{feq}), we can obtain the field equation for $h_{ij}$ as \cite{TZ2019},
\bqn
&&h_{ij}'' + 2 \mathcal{H} h_{ij}'  - \partial^2 h_{ij}  + \frac{\epsilon^{ilk}}{aM_{\rm PV}} \partial_l \Big[ c_1 h_{jk}'' + (\mathcal{H}c_1+c_1') h_{jk}' - c_2 \partial^2 h_{jk}\Big]=0.
\eqn
We find this equation is exactly same with Eq.(3.9) in \cite{TZ2019} and Eq.(3.13) in \cite{Qiao:2019hkz}, which indicates that the field equation in Eq.(\ref{feq}) derived in this article is consistent with the previous works.

\section*{Appendix C: PPN potentials}
\renewcommand{\theequation}{C.\arabic{equation}} \setcounter{equation}{0}

In this Appendix, we present the explicit expressions for the PPN potentials used to parameterize the metric in Eqs. (\ref{h_SP1})-(\ref{h_SP4}). These potentials are given as follows \cite{will2018th}:
\bqn
U & \equiv & \int \frac{\rho\left(\mathbf{x}^{\prime}, t\right)}{\left|\mathbf{x}-\mathbf{x}^{\prime}\right|} d^{3} x^{\prime}, \lb{SPU}\\
\Phi_{i j} & \equiv& \int \frac{\rho\left(\mathbf{x}^{\prime}, t\right)\left(x-x^{\prime}\right)_{i}\left(x-x^{\prime}\right)_{j}}{\left|\mathbf{x}-\mathbf{x}^{\prime}\right|^{3}} d^{3} x^{\prime}, \\
\Phi_{W} & \equiv& \int \rho^{\prime} \rho^{\prime \prime} \frac{\mathbf{x}-\mathbf{x}^{\prime}}{\left|\mathbf{x}-\mathbf{x}^{\prime}\right|^{3}}\left(\frac{\mathbf{x}^{\prime}-\mathbf{x}^{\prime \prime}}{\left|\mathbf{x}-\mathbf{x}^{\prime \prime}\right|}-\frac{\mathbf{x}-\mathbf{x}^{\prime \prime}}{\left|\mathbf{x}^{\prime}-\mathbf{x}^{\prime \prime}\right|}\right) d^{3} x^{\prime} d^{3} x^{\prime \prime} ,\\
\Phi_{1} & \equiv & \int \frac{\rho^{\prime} v^{\prime 2}}{\left|\mathbf{x}-\mathbf{x}^{\prime}\right|} d^{3} x^{\prime} ,\\
\Phi_{2} & \equiv & \int \frac{\rho^{\prime} U^{\prime}}{\left|\mathbf{x}-\mathbf{x}^{\prime}\right|} d^{3} x^{\prime}, \\
\Phi_{3} & \equiv & \int \frac{\rho^{\prime} \Pi^{\prime}}{\left|\mathbf{x}-\mathbf{x}^{\prime}\right|} d^{3} x^{\prime}, \\
\Phi_{4} & \equiv & \int \frac{p^{\prime}}{\left|\mathbf{x}-\mathbf{x}^{\prime}\right|} d^{3} x^{\prime} \\
\mathfrak{A} & \equiv & \int \frac{\rho^{\prime}\left[\mathbf{v}^{\prime} \cdot\left(\mathbf{x}-\mathbf{x}^{\prime}\right)\right]^{2}}{\left|\mathbf{x}-\mathbf{x}^{\prime}\right|^{3}} d^{3} x^{\prime}, \\
\mathfrak{B} & \equiv &\int \rho^{\prime} \mathbf{\nabla}^{\prime} U^{\prime} \cdot   \frac{\left(\mathbf{x}-\mathbf{x}^{\prime}\right)  }{\left|\mathbf{x}-\mathbf{x}^{\prime}\right|} d^{3} x^{\prime},\\
V_{j} & \equiv& \int \frac{\rho\left(\mathbf{x}^{\prime}, t\right) v_{j}^{\prime}}{\left|\mathbf{x}-\mathbf{x}^{\prime}\right|} d^{3} x^{\prime},\\
W_{j} & \equiv &\int \frac{\rho\left(\mathbf{x}^{\prime}, t\right) \mathbf{v}^{\prime} \cdot\left(\mathbf{x}-\mathbf{x}^{\prime}\right)\left(x-x^{\prime}\right)_{j}}{\left|\mathbf{x}-\mathbf{x}^{\prime}\right|^{3}} d^{3} x^{\prime}, \\
\chi & \equiv &-\int \rho\left(\mathbf{x}^{\prime}, t\right)\left|\mathbf{x}-\mathbf{x}^{\prime}\right| d^{3} x^{\prime}.\lb{SPchi}
\eqn

These potentials satisfy the following relations \cite{will2018th},
\bqn
\partial^{2} U &= &-4 \pi G \rho, \\
\nabla^{2} \chi &=& -2 U,  \\
\chi_{, 0 j} &=&V_{j}-W_{j}, \\
\partial^{2} V_{j} &=& -4 \pi G  \rho v_{j}, \\
V_{j, j} &=&-U_{, 0} \\
\partial^{2} \Phi_{1} &=&-4 \pi G \rho v^{2}, \\
 \partial^{2} \Phi_{2} &=&-4 \pi G \rho U=U \partial^{2} U \\
\partial^{2} \Phi_{3} &=&-4 \pi G \rho \Pi, \\
 \partial^{2} \Phi_{4} &=&-4 \pi G  p, \\
2 \chi_{, i j} U_{, i j} &=&\partial^{2}\left(\Phi_{W}+2 U^{2}-3 \Phi_{2}\right), \\
\chi_{, 00} &=&\mathfrak{A}+\mathfrak{B}-\Phi_{1}.
\eqn


\section*{Acknowledgements}

We appreciate the helpful discussion with Mingzhe Li and Dehao Zhao. This work is supported by the National Key Research and Development Program of China Grant under Grant No. 2021YFC2203100 and No.2020YFC2201503, the Zhejiang Provincial Natural Science Foundation of China under Grant No. LR21A050001, the Fundamental Research Funds for the Central Universities under Grant No: WK2030000036 and WK3440000004, the Strategic Priority Research Program of the Chinese Academy of Sciences Grant No. XDB23010200, Key Research Program of the Chinese Academy of Sciences, Grant No. XDPB15, and the science research grants from the China Manned Space Project with NO.CMS-CSST-2021-B01 and CMS-CSST-2021-B11.

\end{document}